\documentclass[reprint,twocolumn,showpacs,preprintnumbers,amsmath,amssymb,aps,prc,]{revtex4-1}

\usepackage{graphicx}
\usepackage{dcolumn}
\usepackage{bm}
\usepackage{lineno}
\usepackage{float}
\usepackage{upgreek}
\usepackage{booktabs}
\usepackage{siunitx}
\usepackage{graphicx} 
\usepackage{subfigure}

\begin{document}

\title{The optimal detection angles for producing N=126 neutron-rich isotones in the multinucleon transfer reactions}


\author{Zehong Liao$^{1}$}
\author{Long Zhu$^{1}$}
\thanks{Corresponding author: zhulong@mail.sysu.edu.cn}
\author{Zepeng Gao$^{1}$}
\author{Jun Su$^{1}$}
\author{Cheng Li$^{2}$}

\affiliation{%
$^{1}$Sino-French Institute of Nuclear Engineering and Technology, Sun Yat-sen University, Zhuhai 519082, China\\
$^{2}$College of Physics and Technology and Guangxi Key Laboratory of Nuclear Physics and Technology, Guangxi Normal University, Guilin 541004, China\\
}%

\date{\today}

\begin{abstract}

The challenge of isotopic identification over a wide angular distribution has limited the measurement of neutron-rich nuclei produced via the multinucleon transfer (MNT) process. To investigate the optimal detection angles for the N=126 isotones, we propose a method to construct the reasonable scattering angles of the MNT products in the dinuclear system (DNS-sysu) model. The reactions $^{136,144}$Xe + $^{208}$Pb are investigated. The calculated results are in rather good agreement with the available experimental data in the reaction $^{136}$Xe + $^{208}$Pb. The entrance channel effects on the scattering angle are investigated. It is found that the scattering angular distribution strongly depends on the isospin and the impact parameter of the collision system. The optimal angle ranges for detecting $N=126$ neutron-rich nuclides $^{204}$Pt, $^{203}$Ir, $^{202}$Os, and $^{201}$Re in the $^{136}$Xe + $^{208}$Pb reaction at the incident energy $\mathrm{E_{c.m.} = \SI{526}{MeV}}$ are predicted. Our results suggest that the angle range $45^{\circ} \leqslant \theta_{\mathrm{lab}} \leqslant 50^{\circ}$ is most favorable for detecting unknown N=126 isotones. Given the current difficulties in separating and identifying experimental MNT fragments, the results of this work could provide significant contributions to future experiments.
\end{abstract}


\maketitle

$Introduction.$ To date, a great deal of effort has been focused on the production of neutron-rich nuclei around $N=126$, which not only sheds light on exotic nuclei properties but also provides crucial insights into astrophysically important processes \cite{Grawe2007}. The remarkable recent progress in the synthesis of neutron-rich nuclei has been made via fusion, fission, and fragmentation. Nevertheless, the capabilities of these methods have increasingly weakened when attempting to produce neutron-rich nuclei near $N=126$ and beyond. Thus, the dilemma of the synthesis of new neutron-rich nuclei arouses the expectation of the advent of an alternative approach \cite{Volkov1978, Zagrebaev2008, Corradi2013, Zhang2018, Adamian2020, Saiko2022}.

Thank to the recent development in direct isotopic identification, in particular, by using large acceptance magnetic spectrometers for heavy ion reactions, the promised potential of the multinucleon transfer (MNT) reaction was established for the first time in the collision of $^{136}$Xe + $^{198}$Pt \cite{Watanabe2015}. Consequently, the MNT reactions have gained renewed interest in terms of the production of new isotopes around and beyond the neutron shell $N=126$ experimentally \cite{Kozulin2012, Barrett2015, Vogt2015, Kozulin2017, Diklic2023}. However, the lack of sufficiently sensitive identification techniques for MNT products is still a serious bottleneck for further producing unknown nuclides. One of the major challenges in detecting unknown isotopes is the non-isotropic angular distribution \cite{Wilczy1973} produced in the MNT reaction. In the case of fusion evaporation and fragmentation reactions, products are typically emitted at a narrow forward angle around $0^{\circ}$ in the laboratory system, whereas MNT products cover a wide cone angle. In addition, the covered angle varies with the reaction system and reaction products, making them less effective for collection and separation \cite{Heinz2022, Valverde2020}. 

Theoretically, various phenomenological or quantum microscopic approaches including the multidimensional Langevin model \cite{Karpov2017, Saiko2019}, the dinuclear system (DNS) model \cite{Feng2017, Bao2018, Zhu2018, Guo2019}, the improved quantum molecular dynamics model (ImQMD) \cite{Li2019, Zhao2021, Zhao2022}, Time-dependent Hartree-Fock (TDHF) theory \cite{Sekizawa2016, Jiang2018, Guo2019, Sun2023}, and the stochastic mean-field approach (SMF) \cite{Ayik2021, Ayik2023} have been proposed to study the MNT reaction mechanism in low-energy heavy-ion collisions. Fruitful works have been done systematically to find the best reaction conditions by manipulating the parameters of the collision entrance channel (projectile-target combinations and energy). However, the theoretical prediction and guidance on the most likely emission angle of nuclides are scarce. Developing a comprehensive and accurate description of the MNT dynamics to optimize detection efficiency is essential.

Originating from the DNS concept proposed by Volkov in the deep inelastic collision (DIC) \cite{Volkov1978}, the DNS model is gradually developed and can be applied to describe multiple reaction channels including quasifission, fusion, and multinucleon processes \cite{Adamian1997, Li2003, Feng2007, Zhu2021unified}. While the DNS model is quite successful in reproducing the probability distribution of collective variables including the mass and charge yield, it fails to provide reasonable fragment information related to the final scattering angle. Here, we propose a method to construct the reasonable scattering angles of the MNT products in the DNS-sysu model. This extension also provides a theoretical basis for subsequent conclusions in this letter. 

\begin{figure*}
    \centering
        \includegraphics[width=16cm]{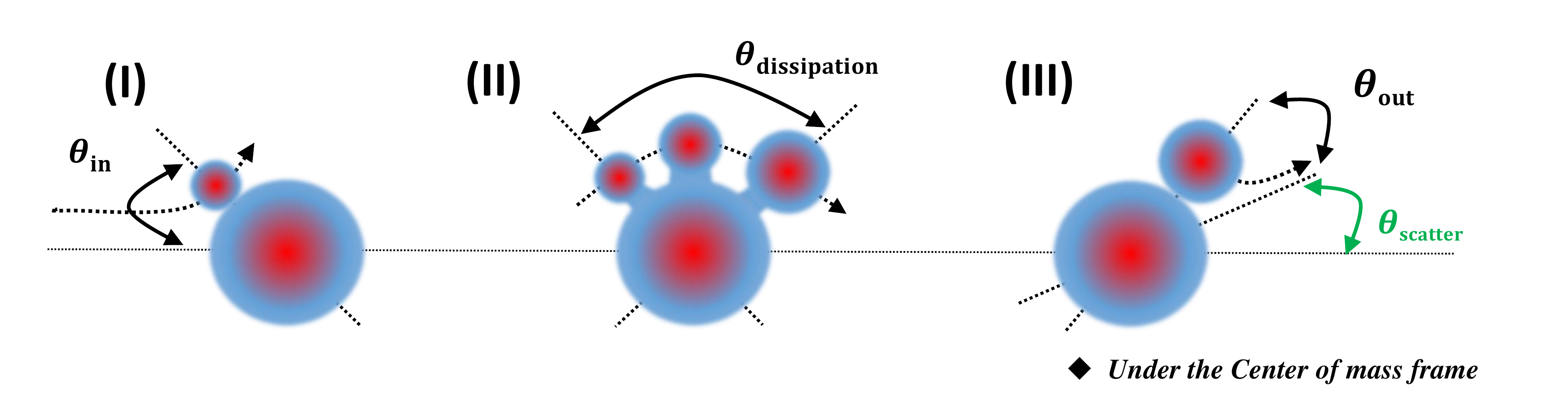}
    \caption{The schematic diagram of the evolution process of a dinuclear system. The dotted arrows and solid arrows respectively represent the motion trajectories and rotation angles at each stage. } \label{figure1}
\end{figure*}

$\emph{Probability distribution in the DNS-sysu model.}$ 
The master equation is one of the most suitable mathematical tools to describe non-equilibrium statistical processes. In the DNS-sysu model, the fragment distribution probability $P$ can be calculated by numerically solving the following master equation:
\begin{flalign}
\begin{split}\label{master}
&\frac{dP(Z_{1}, N_{1}, \beta_{2}, J, t)}{dt}\\
&=\sum_{Z_{1}^{'}}W_{Z_{1}, N_{1}, \beta_{2}; Z_{1}^{'}, N_{1}, \beta_{2}}(t)[d_{Z_{1}, N_{1}, \beta_{2}}P(Z_{1}^{'}, N_{1}, \beta_{2}, J, t)\\
&-d_{Z_{1}^{'}, N_{1}, \beta_{2}}P(Z_{1}, N_{1}, \beta_{2}, J, t)]\\
&+\sum_{N_{1}^{'}}W_{Z_{1}, N_{1}, \beta_{2};Z_{1}, N_{1}^{'}, \beta_{2}}(t)[d_{Z_{1}, N_{1}, \beta_{2}}P(Z_{1}, N_{1}^{'}, \beta_{2}, J, t)\\
&-d_{Z_{1}, N_{1}^{'}, \beta_{2}}P(Z_{1}, N_{1}, \beta_{2}, J, t)]\\
&+\sum_{\beta_{2}^{'}}W_{Z_{1}, N_{1}, \beta_{2};Z_{1}, N_{1}, \beta_{2}^{'}}(t)[d_{Z_{1}, N_{1}, \beta_{2}}P(Z_{1}, N_{1}, \beta_{2}^{'}, J, t)\\
&-d_{Z_{1}, N_{1}, \beta_{2}^{'}}P(Z_{1}, N_{1}, \beta_{2}, J, t)].
\end{split}
\end{flalign}
Here, $W_{Z_{1}, N_{1}, \beta_{2};Z_{1}^{'}, N_{1}, \beta_{2}}$ denotes the mean transition probability from the channel ($Z_{1}$, $N_{1}$, $\beta_{2}$) to ($Z_{1}^{'}$, $N_{1}$, $\beta_{2}$), which is similar to $N_{1}$ and $\beta_{2}$. $d_{Z_{1}, N_{1}, \beta_{2}}$ is the microscopic dimension (the number of channels) corresponding to the macroscopic state ($Z_{1}$, $N_{1}$, $\beta_{2}$). For the degrees of freedom of charge and neutron number, the sum is taken over all possible proton and neutron numbers that fragment 1 may take, but only one nucleon transfer is considered in the model ($Z_{1}^{'} = Z_{1} \pm 1$; $N_{1}^{'} = N_{1} \pm 1$). For the $\beta_{2}$, we take the range of -0.5 to 0.5. The evolution step length is 0.01. The transition probability is related to the local excitation energy, in which the memory time is $0.25\tau_{\textrm{0}}/\mathit{A}$. Here, $\tau_{0} \equiv 2\pi \hbar/(1 \textrm{MeV})\approx 4\times 10^{-21}$ \textrm{s} , and $\mathit{A}$ means the total nucleon number of the reaction.

In our previous work \cite{Zhu2021, Zhu2022, Liao2023}, the developed model was successfully applied to the analysis of production cross sections of neutron-rich heavy nuclei in MNT reactions, and a more detailed description of the model can be seen in Ref. \cite{Zhu2021unified}. In this work, we propose a method for calculating the angular distribution of the MNT products in the DNS-sysu model 


$\emph{Scattering angle in the DNS-sysu model.}$ In Fig. \ref{figure1}, we present a diagram of the evolution process during the collision to illustrate the formation of the scattering angle. According to the diagram: (\uppercase\expandafter{\romannumeral1}) The projectile follows the Coulomb trajectory and sticks into a DNS with the target nuclei. (\uppercase\expandafter{\romannumeral2}) The system rotates at a certain angle $\theta _{\mathrm{dissipation}}$, accompanied by energy dissipation and exchange of nucleons. (\uppercase\expandafter{\romannumeral3}) The separated binary products move to infinity along the Coulomb trajectories. Eventually, the observed scattering angle of the projectile-like fragment (PLF) can be measured by the given relation \cite{Wolschin1978, Riedel1979, Toke1985}:
\begin{align}
\theta_{\mathrm{scatter}} = \pi - \theta_{\mathrm{in}} - \theta_{\mathrm{dissipation}} - \theta_{\mathrm{out}},
\end{align}
where the ingoing Coulomb deflection angle $\theta_{\mathrm{in}}$ and the outgoing deflection angle $\theta _{\mathrm{out}}$ are determined by the Coulomb trajectories in entrance and exit channels with the corresponding values for energy $E$ and impact parameter $b$.
\begin{align}
\theta_{\mathrm{in(out)}} =  \arcsin \frac{2b/R+\varepsilon}{\sqrt{4+\varepsilon^{2}}} - \arcsin \frac{1}{\sqrt{(2/\varepsilon)^2}+1}.
\end{align}
Here, $\varepsilon=Z_{1}Z_{2}e^{2}/(Eb)$ and $R$ means the interaction radius, which is related to the experimental cross section in the classical approximation \cite{Wolschin1978}. Subsequently, the evaluation of the $\theta _{\mathrm{dissipation}}$ is involving the estimation of the sticking time \cite{Li1983} and moment of inertia of the DNS. It is assumed that the two colliding nuclei contact and stick at $R_{\mathrm{cont}}$ and the angular velocity is time-dependent. During the stick time $t_{s}$ between contact and scission, $\theta _{\mathrm{dissipation}}$ is calculated by
\begin{align}
\theta_{\mathrm{dissipation}} =  \omega _{\mathrm{DNS}} * t_{s} =\int_{}^{t_{s} } \frac{J(t)\hbar }{\left \langle I  \right \rangle } \mathrm{d}t  .
\end{align}
Here, $J(t) [=J_{st}+(J-J_{st}e^{\frac{-t}{\tau _{j} } }  ) ]$ is the angular momentum at time t. To simplify the calculations, we have considered only the mean moment of inertia $\left \langle I  \right \rangle$ with the rigid-body approximation \cite{Shen1987}. For the relative motion of the dinuclear system, the mean moment of inertia  $\left \langle I  \right \rangle$ can be estimated as the average of the entrance and exit moments of inertia:
\begin{align}
\left \langle I  \right \rangle = \frac{1}{2}*(I_{\mathrm{in}}+I_{\mathrm{out}}),
\end{align}
where the entrance and exit moment of inertia $I$ is determined by the entrance and exit channels, respectively, $e.g.,$ the corresponding values for the reduced mass of the composite system $\mu$, the position where the nucleon transfer process takes place $R_{\mathrm{cont}}$ in the entrance channel (exit channel). $R_{\mathrm{cont}}$ is calculated as $R_{1} + R_{2} + 0.7$ fm for consistency with the calculation of the potential energy surface \cite{Zhu2021}. 

The differential angular distribution for the primary fragment with charge number $Z_{1}$, neutron number $N_{1}$, and scattering angle $\theta$ in the MNT reaction can be calculated as:
\begin{flalign}
  \begin{split}
 &\frac{d^{3}\sigma_{\textrm{pr}}(Z_{1},N_{1},\theta)}{dZ_{1}dN_{1}d\Omega } = \frac{\pi\hbar^{2}}{2\mu E_{\textrm{c.m.}}}\\
 &\times\int_{J = 0}^{J_{\textrm{max}}}(2J+1)T_{\textrm{cap}}(J)\frac{\Delta P(Z_{1},N_{1}, \theta,J)\mathrm{d}J}{ 2 \pi \sin \theta \Delta Z_{1}\Delta N_{1} \Delta \theta },
 \end{split}
\end{flalign}
 where $P(Z_{1}, N_{1}, \theta, J)$ is the probability distribution of $Z_{1}$, $N_{1}$, and $\theta$ at the initial angular momentum $J$, which is determined by solving the master Eq. (\ref{master}). Here, the distribution of scattering angle $\theta$ at the initial angular momentum $J$ is resulted from the $\beta_{2}$ distribution by affecting the moment of inertia of the DNS. In other words, the DNS-sysu model constructs $\theta_{\mathrm{scatter}}$ fluctuation  for generating a specific fragment $(Z_{1}, N_{1})$ with an initial angular momentum of $J$ by introducing the $\beta_{2}$ degree of freedom. 

$\emph{Results and discussions.}$ To verify the above method, the experimental data \cite{Kozulin2012} for $^{136}$Xe + $^{208}$Pb collisions at two incident energies $E_{c.m.}=526$ and \SI{617}{MeV} have been analyzed. Note that some restrictions are imposed in this experiment. The coverage angular range of $25^{\circ} < \theta_{\mathrm{lab}} < 70^{\circ}$ and total kinetic energy losses larger than \SI{40}{MeV} are taken into account in the following calculations as well.

\begin{figure}[]
\centering
\includegraphics[width=8.5cm]{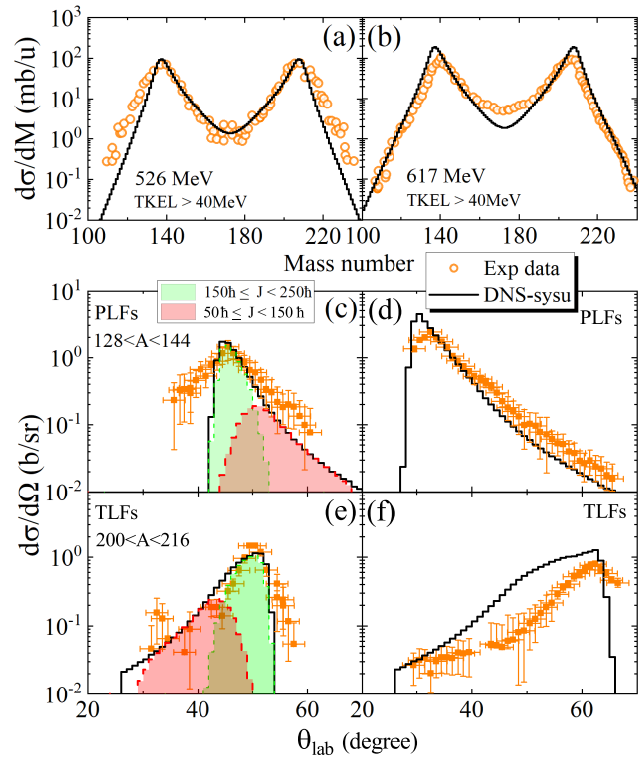}
\caption{Mass and angular distributions of primary fragments in the $^{136}$Xe + $^{208}$Pb reaction for two collision energies $\mathrm{E_{c.m.} = \SI{526}{MeV}}$ [panels (a), (c), (e)] and $\mathrm{E_{c.m.} = \SI{617}{MeV}}$ [panels (b), (d), (f)]. The experimental data points (symbols) are from Ref. \cite{Kozulin2012}. In order to obtain the same normalization as the model results, the original experimental data of the mass distribution is multiplied by two times. The angular distributions are shown for light reaction fragments with $\mathrm{128 < A < 144}$ and heavy fragments with $\mathrm{200 < A < 216}$.}\label{figure2}
\end{figure}

 The theoretical and experimental distributions of binary primary products are compared in Fig. \ref{figure2}. For the distribution at the energy $\mathrm{E_{c.m.} = \SI{526}{MeV}}$ shown in the left panels, one can see that the calculation results (denoted with thick solid black lines) are in good agreement with the corresponding experimental data (denoted with symbols) for both mass and angular distribution, except for a slightly narrower theoretical angular distribution compared to the experimental results in Fig. \ref{figure2}(c). In addition, the angular distributions in the entrance angular momentum ranges of $50 \hbar \leqslant J < 150 \hbar$ and $150 \hbar \leqslant J < 250 \hbar$ are also shown, revealing a broader distribution of PLFs for $50 \hbar \leqslant J < 150 \hbar$ compared to peripheral collisions ($e.g., 150 \hbar \leqslant J < 250 \hbar$). As the angular momentum $J$ decreases, the strong correlation between the rotation angle and interaction time widens and extends the distribution of TLFs towards the forward angle region. For the collision with a higher incident energy $\mathrm{E_{c.m.} = \SI{617}{MeV}}$ of the right panels (b), (d), and (f), the peak position and the shape of the distribution are also in good agreement with the experimental data. Since the Coulomb trajectory is always influenced by the incident energy, the enhanced Coulomb deflection of the DNS results in PLFs being emitted at more forward angles with respect to the beam direction at higher incident energy. We notice that the calculated result slightly overestimates the experimental data for TLFs angular distribution at $\mathrm{E_{c.m.} = \SI{617}{MeV}}$ as shown in Fig. \ref{figure2}(f). In the future, introducing more degrees of freedom would make better the description of the experimental data by the DNS-sysu model. 
 
\begin{figure}[]
\centering
\includegraphics[width=8.5cm]{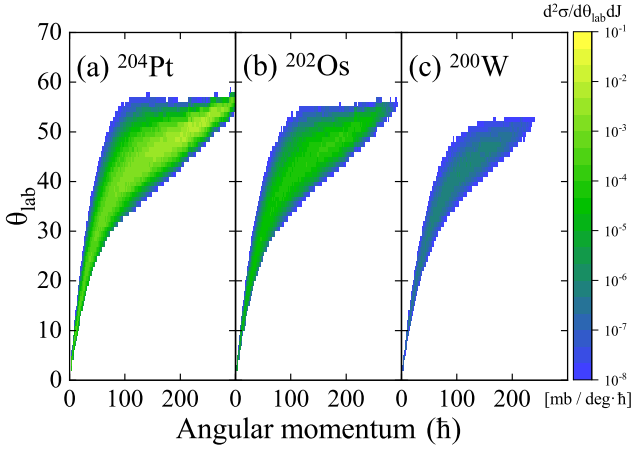}
\caption{The double differential cross section distributions d$^{2}\sigma$/d$\theta_{\mathrm{lab}}$d$J$ of (a) $^{204}$Pt, (b) $^{202}$Os, and (c) $^{200}$W in the reaction $^{136}$Xe + $^{208}$Pb at $\mathrm{E_{c.m.} = \SI{526}{MeV}}$.} \label{pic-momentum}
\end{figure}

As mentioned above, the angular distribution strongly depends on the entrance angular momentum. Considering the great scientific interests, we show the correlation between the final scattering angle of the $N=126$ isotones and the entrance angular momentum. The double differential cross section distributions d$^{2}\sigma$/d$\theta_{\mathrm{lab}}$d$J$ of $^{204}$Pt, $^{202}$Os, and $^{200}$W in the reaction $^{136}$Xe + $^{208}$Pb at $\mathrm{E_{c.m.} = \SI{526}{MeV}}$ are shown in Fig. \ref{pic-momentum}. For each isotone, we can notice that the width of the scattering angular distribution initially broadens with decreasing angular momentum from grazing collisions. The most probable production of neutron-rich nuclides occurs within the angular momentum range of ($100\hbar < J < 200\hbar$). As the collisions become more violent ($J < 100\hbar$), the angular distribution of the reaction products narrows and shifts towards more forward angles. Even for central collisions ($J \approx 0\hbar$), $\theta_{c.m.} \approx 0 ^{\circ}$ is observed. This is because the strong Coulomb repulsion pushes the nuclei apart and causes the TLFs to be emitted mostly in the forward direction, with a narrow angular distribution around the beam axis. In addition to the exponential decline of the absolute differential cross section, the narrowing and isotropization of the TLF angle distribution can be observed as the atomic number decreases ($\Delta Z=- 4, - 6, - 8 $).

\begin{figure}[]
\centering
\includegraphics[width=8.5cm]{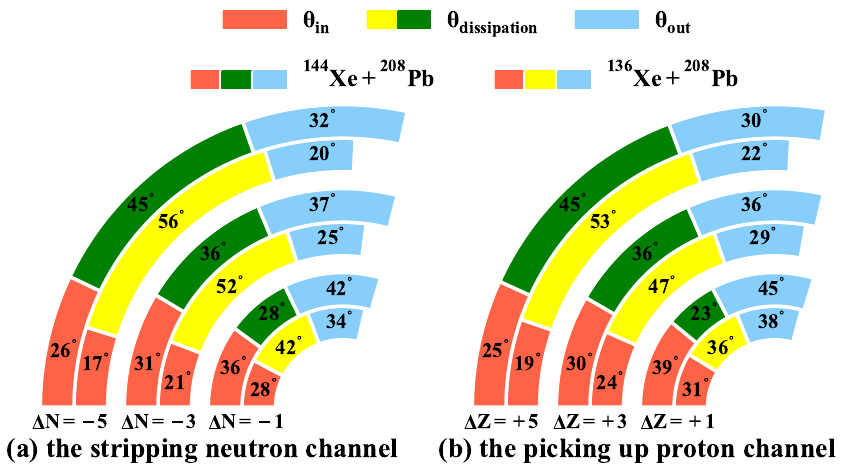}
\caption{Left panel: A pattern composed of angles $\theta_{\mathrm{in}}$, $\theta_{\mathrm{dissipation}}$, and $\theta_{\mathrm{out}}$ in the neutron stripping channel for the reactions $^{136,144}$Xe + $^{208}$Pb. The angles $\theta_{\mathrm{in}}$ and $\theta_{\mathrm{out}}$ are denoted with red and blue patterns, respectively. The green and yellow patterns denote the angles of $\theta_{\mathrm{dissipation}}$ in the reactions induced by $^{144}$Xe and $^{136}$Xe, respectively. Right panel: the same as the left panel but for proton picking up channels.} \label{pic-isospin}
\end{figure}

Clearly, the shape of the angular distribution is closely related to not only the entrance angular momentum but also the nucleon transfer channel. To gain insight into the production mechanisms and characteristics of the neutron-rich fragments from the viewpoint of the angular distribution, we investigate in detail by comparing the angles of the three parts ($\theta_{\mathrm{in}}$, $\theta_{\mathrm{dissipation}}$, and $\theta_{\mathrm{out}}$) for different nucleon transfer channels in the reactions $^{136, 144}$Xe + $^{208}$Pb. In the left panels of Fig. \ref{pic-isospin}, we present the average values of $\theta_{\mathrm{in}}$, $\theta_{\mathrm{dissipation}}$, and $\theta_{\mathrm{out}}$ corresponding to the PLFs in the reactions $^{136, 144}$Xe + $^{208}$Pb for the neutron stripping channels ($\Delta N =$ -1, -3, and -5, respectively). The incident energy for both reactions is $\mathrm{E_{c.m.} = \SI{526}{MeV}}$. Note that the angles are given in the center of the mass frame and obtained by weighting the fragment probabilities. The contribution of each part, as shown in Fig. \ref{pic-isospin}, is also numerically shown. One can see that the discrepancy in the final scattering angle between the two reactions increases and the proportion of $\theta_{\mathrm{in}}$, $\theta_{\mathrm{dissipation}}$, and $\theta_{\mathrm{out}}$ has a significant change with the increasing number of the transferred neutron. In the -1n transfer channel, it is evident that the Coulomb deflection angle in the reaction $^{144}$Xe + $^{208}$Pb is larger compared to that in the $^{136}$Xe + $^{208}$Pb reaction. This is because the radius of $^{144}$Xe projectile is larger, and the magnitude of the Coulomb deflection angle is influenced by the impact parameters related to the radius. As the average results shown in Fig. \ref{pic-isospin}, the more transferred nucleons also imply that the production of fragments mainly occurs in more violent collisions, $i.e.$ lower impact parameter. As a consequence, the Coulomb deflection angle decreases gradually ($e.g., 28 ^{\circ} (34 ^{\circ}) \rightarrow 21 ^{\circ} (25 ^{\circ}) \rightarrow 17 ^{\circ} (20 ^{\circ})$ for $\theta_{\mathrm{in(out)}}$ in $^{136}$Xe + $^{208}$Pb).

Another important feature is the discrepancy of the $\theta_{\mathrm{dissipation}}$ in the reactions $^{136,144}$Xe + $^{208}$Pb. Unlike the Coulomb deflection angle, we notice that the values of $\theta_{\mathrm{dissipation}}$ in the reaction $^{144}$Xe + $^{208}$Pb are slightly smaller than those in $^{136}$Xe + $^{208}$Pb, $e.g.,$ $28^{\circ}<42^{\circ}$, $36^{\circ}<52^{\circ}$, and $45^{\circ}<56^{\circ}$ for the channels $\Delta N =$ -1, -3, and -5, respectively. These behaviors can be interpreted as the results of charge equilibration: the N/Z values of $^{136}$Xe, $^{144}$Xe, and $^{208}$Pb is 1.52, 1.66, and 1.54, respectively. In the dynamical neutron transfer process, $^{144}$Xe is more inclined to lose neutrons in a short period of time compared to $^{136}$Xe, which results in a smaller rotation angle $\theta_{\mathrm{dissipation}}$ in the $^{144}$Xe induce reaction. For the proton pickup channels ($\Delta Z =$ +1, +3, and +5, respectively), as shown in Fig. \ref{pic-isospin}(b), a similar behavior is presented due to the similar charge equilibration effects.

\begin{figure}
\centering
\includegraphics[width=8cm]{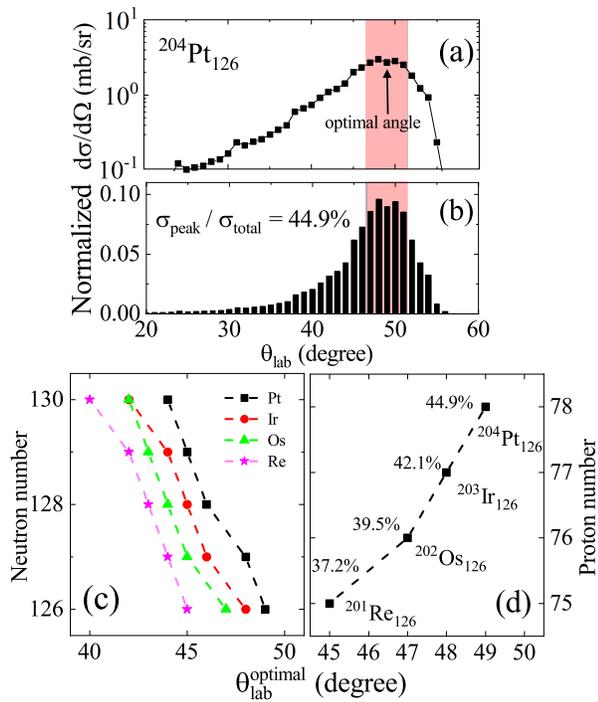}
\caption{The calculated angular distribution (a) and the corresponding normalized one (b) of $^{204}$Pt fragments produced in the $^{136}$Xe+$^{208}$Pb system at $\mathrm{E_{c.m.} = \SI{526}{MeV}}$. The arrow indicates the optimal detected angle of the $^{204}$Pt isotope. (c) The optimal detected angles of unknown Pt, Ir, Os, and Re isotopes in the laboratory system in the $^{136}$Xe+$^{208}$Pb at $\mathrm{E_{c.m.} = \SI{526}{MeV}}$. The ratios of production yields of N=126 isotones emitted within the range of 5 degrees around the optimal detection angle are denoted in (d).} \label{pic-bestangle}
\end{figure}

The dependence of the scattering angle on the nucleon transfer channel could cause the discrepancy of optimal detection angles for different $N=126$ isotones. Also, considering the facts that one of the biggest difficulties in the production of the neutron-rich unknown isotopes is the low efficiencies of separation and detection in the experiments, we extract the peak locations in the angular distributions, $i.e.$ the optimal detection angles where the objective isotopes are most likely to be produced. Within the DNS-sysu model, the angular distribution of $^{204}$Pt in the reaction $^{136}$Xe+$^{208}$Pb at $\mathrm{E_{c.m.} = \SI{526}{MeV}}$ is shown in Fig. \ref{pic-bestangle}(a). The peak feature of the distribution is clearly presented. We locate the most probable scattering angle by the total cross section within the range of 5 degrees ($\sum_{\theta -2}^{\theta +2} 2\pi \sigma (\theta )\sin \theta$) in the laboratory system, which is denoted with the red-shaded area. It can be seen that $^{204}$Pt could be produced most likely at the backward angle of $49^{\circ}$. To evaluate the ratio of yields in the optimal detection angle range, the normalized distribution is shown in Fig. \ref{pic-bestangle}(b). We find that the ratio of 44.9 $\%$ production yields can be detected in the angle range of $47^{\circ} \leqslant \theta_{\mathrm{lab}} \leqslant 51^{\circ}$. Close to half of the yields could be detected in a small range of angles around the optimal one. Therefore, the investigation of the most probable scattering angle for producing each $N=126$ isotone is necessary.

In Fig. \ref{pic-bestangle} (c), the optimal detection angles of the Pt, Ir, Os, and Re isotopes produced in the reaction $^{136}$Xe+$^{208}$Pb at $\mathrm{E_{c.m.} = \SI{526}{MeV}}$ are displayed. Note that all isotopic chains exhibit similar behavior, $i.e.$, the optimal detection angles shift towards the forward angles as the neutron number of the isotopes increases. It indicates that the neutron-rich nuclides are more favorable to be produced at the forward angle region. In addition, we also extracted the optimal detection angles and the corresponding cross section proportion for all four $N=126$ neutron-rich isotones ($^{204}$Pt, $^{203}$Ir, $^{202}$Os, and $^{201}$Re) and shown in Fig. \ref{pic-bestangle}(d). In particular, the cross section proportions ($44.9 \% \rightarrow 42.1 \% \rightarrow 39.5 \% \rightarrow 37.2 \%$ ) decrease for producing more neutron-rich isotones, which is associated with the violent collisions and long interaction times taking place at small impact parameters. The similar behavior was noticed recently in Ref. \cite{Karpov2017} within the multidimensional Langevin model. The optimal detection angles of the four $N=126$ neutron-rich nuclides $^{204}$Pt, $^{203}$Ir, $^{202}$Os, and $^{201}$Re are around $45^{\circ} \leqslant \theta_{\mathrm{lab}} \leqslant 50^{\circ}$ in the $^{136}$Xe + $^{208}$Pb at $\mathrm{E_{c.m.} = \SI{526}{MeV}}$. This provides an important basis for further experiments on the synthesis of neutron-rich nuclei around $N=126$ in MNT reactions.

$\emph{Conclusions.}$ We propose a method for calculating the angular distribution of the MNT products. The reasonable description of angular distribution is realized for the first time based on the framework of the DNS model. The calculated results are in rather good agreement with the available measurements in the reaction $^{136}$Xe + $^{208}$Pb. Furthermore, we carry out an investigation of entrance channel effects on the angular distribution in the reactions $^{136,144}$Xe + $^{208}$Pb. The dependence of the isospin on the final scattering angle is found in the reactions $^{136,144}$Xe + $^{208}$Pb. Compared to the $^{136}$Xe induced reaction, it is noticed that the PLF produced in the collision of $^{144}$Xe + $^{208}$Pb tend to the forward angle, and the contributions of $\theta_{\mathrm{in}}$, $\theta_{\mathrm{dissipation}}$, and $\theta_{\mathrm{out}}$ to the total final scattering angle have significant variances. The obvious effect of entrance angular momentum on the scattering angle distribution is also noticed in the reaction $^{136}$Xe + $^{208}$Pb. From the landscape of the double differential cross sections d$^{2}\sigma$/d$\theta_{\mathrm{lab}}$d$J$, a more isotropic angle distribution can be seen with the decrease of atomic number ($\Delta Z=- 4, - 6, - 8 $). And, the underlying mechanisms are analyzed according to the quantitative calculation. 

The optimal detection angles of the Pt, Ir, Os, and Re isotopes produced in the $^{136}$Xe+$^{208}$Pb reaction have been investigated. It is found that (i) the neutron-rich nuclides are favorable to be produced at the forward angle region and (ii) exceeding 40$\%$ of total yields of several N=126 unknown isotones produced in the reaction $^{136}$Xe+$^{208}$Pb could be detected within an angle range of 5$^{\circ}$ around the optimal angles. Finally, we predict that $45^{\circ} \leqslant \theta_{\mathrm{lab}} \leqslant 50^{\circ}$ is the most favorable angular range to detect neutron-rich nuclei around $N=126$.

$\emph{Acknowledgments.}$ This work was supported by the National Natural Science Foundation of China under Grants No. 12075327 and 11805015; Guangdong Major Project of Basic and Applied Basic Research under Grant No. 2021B0301030006.
\bibliography{reference}

\begin{thebibliography}{42}%
\makeatletter
\providecommand \@ifxundefined [1]{%
 \@ifx{#1\undefined}
}%
\providecommand \@ifnum [1]{%
 \ifnum #1\expandafter \@firstoftwo
 \else \expandafter \@secondoftwo
 \fi
}%
\providecommand \@ifx [1]{%
 \ifx #1\expandafter \@firstoftwo
 \else \expandafter \@secondoftwo
 \fi
}%
\providecommand \natexlab [1]{#1}%
\providecommand \enquote  [1]{``#1''}%
\providecommand \bibnamefont  [1]{#1}%
\providecommand \bibfnamefont [1]{#1}%
\providecommand \citenamefont [1]{#1}%
\providecommand \href@noop [0]{\@secondoftwo}%
\providecommand \href [0]{\begingroup \@sanitize@url \@href}%
\providecommand \@href[1]{\@@startlink{#1}\@@href}%
\providecommand \@@href[1]{\endgroup#1\@@endlink}%
\providecommand \@sanitize@url [0]{\catcode `\\12\catcode `\$12\catcode
  `\&12\catcode `\#12\catcode `\^12\catcode `\_12\catcode `\%12\relax}%
\providecommand \@@startlink[1]{}%
\providecommand \@@endlink[0]{}%
\providecommand \url  [0]{\begingroup\@sanitize@url \@url }%
\providecommand \@url [1]{\endgroup\@href {#1}{\urlprefix }}%
\providecommand \urlprefix  [0]{URL }%
\providecommand \Eprint [0]{\href }%
\providecommand \doibase [0]{http://dx.doi.org/}%
\providecommand \selectlanguage [0]{\@gobble}%
\providecommand \bibinfo  [0]{\@secondoftwo}%
\providecommand \bibfield  [0]{\@secondoftwo}%
\providecommand \translation [1]{[#1]}%
\providecommand \BibitemOpen [0]{}%
\providecommand \bibitemStop [0]{}%
\providecommand \bibitemNoStop [0]{.\EOS\space}%
\providecommand \EOS [0]{\spacefactor3000\relax}%
\providecommand \BibitemShut  [1]{\csname bibitem#1\endcsname}%
\let\auto@bib@innerbib\@empty
\bibitem [{\citenamefont {Grawe}\ \emph {et~al.}(2007)\citenamefont {Grawe},
  \citenamefont {Langanke},\ and\ \citenamefont
  {Martínez-Pinedo}}]{Grawe2007}%
  \BibitemOpen
  \bibfield  {author} {\bibinfo {author} {\bibfnamefont {H.}~\bibnamefont
  {Grawe}}, \bibinfo {author} {\bibfnamefont {K.}~\bibnamefont {Langanke}}, \
  and\ \bibinfo {author} {\bibfnamefont {G.}~\bibnamefont {Martínez-Pinedo}},\
  }\bibinfo {title} {Nuclear structure and astrophysics},\ \href {\doibase
  10.1088/0034-4885/70/9/R02} {\bibfield  {journal} {\bibinfo  {journal}
  {Reports on Progress in Physics}\ }\textbf {\bibinfo {volume} {70}},\
  \bibinfo {pages} {1525} (\bibinfo {year} {2007})}\BibitemShut {NoStop}%
\bibitem [{\citenamefont {Volkov}(1978)}]{Volkov1978}%
  \BibitemOpen
  \bibfield  {author} {\bibinfo {author} {\bibfnamefont {V.}~\bibnamefont
  {Volkov}},\ }\bibinfo {title} {Deep inelastic transfer reactions — The new
  type of reactions between complex nuclei},\ \href {\doibase
  10.1016/0370-1573(78)90200-4} {\bibfield  {journal} {\bibinfo  {journal}
  {Physics Reports}\ }\textbf {\bibinfo {volume} {44}},\ \bibinfo {pages} {93}
  (\bibinfo {year} {1978})}\BibitemShut {NoStop}%
\bibitem [{\citenamefont {Zagrebaev}\ and\ \citenamefont
  {Greiner}(2008)}]{Zagrebaev2008}%
  \BibitemOpen
  \bibfield  {author} {\bibinfo {author} {\bibfnamefont {V.}~\bibnamefont
  {Zagrebaev}}\ and\ \bibinfo {author} {\bibfnamefont {W.}~\bibnamefont
  {Greiner}},\ }\bibinfo {title} {Production of New Heavy Isotopes in
  Low-Energy Multinucleon Transfer Reactions},\ \href {\doibase
  10.1103/PhysRevLett.101.122701} {\bibfield  {journal} {\bibinfo  {journal}
  {Physical Review Letters}\ }\textbf {\bibinfo {volume} {101}},\ \bibinfo
  {pages} {122701} (\bibinfo {year} {2008})}\BibitemShut {NoStop}%
\bibitem [{\citenamefont {Corradi}\ \emph {et~al.}(2013)\citenamefont
  {Corradi}, \citenamefont {Szilner}, \citenamefont {Pollarolo}, \citenamefont
  {Montanari}, \citenamefont {Fioretto}, \citenamefont {Stefanini},
  \citenamefont {Valiente-Dobón}, \citenamefont {Farnea}, \citenamefont
  {Michelagnoli}, \citenamefont {Montagnoli}, \citenamefont {Scarlassara},
  \citenamefont {Ur}, \citenamefont {Mijatović}, \citenamefont {Malenica},
  \citenamefont {Soić},\ and\ \citenamefont {Haas}}]{Corradi2013}%
  \BibitemOpen
  \bibfield  {author} {\bibinfo {author} {\bibfnamefont {L.}~\bibnamefont
  {Corradi}}, \bibinfo {author} {\bibfnamefont {S.}~\bibnamefont {Szilner}},
  \bibinfo {author} {\bibfnamefont {G.}~\bibnamefont {Pollarolo}}, \bibinfo
  {author} {\bibfnamefont {D.}~\bibnamefont {Montanari}}, \bibinfo {author}
  {\bibfnamefont {E.}~\bibnamefont {Fioretto}}, \bibinfo {author}
  {\bibfnamefont {A.}~\bibnamefont {Stefanini}}, \bibinfo {author}
  {\bibfnamefont {J.}~\bibnamefont {Valiente-Dobón}}, \bibinfo {author}
  {\bibfnamefont {E.}~\bibnamefont {Farnea}}, \bibinfo {author} {\bibfnamefont
  {C.}~\bibnamefont {Michelagnoli}}, \bibinfo {author} {\bibfnamefont
  {G.}~\bibnamefont {Montagnoli}}, \bibinfo {author} {\bibfnamefont
  {F.}~\bibnamefont {Scarlassara}}, \bibinfo {author} {\bibfnamefont
  {C.}~\bibnamefont {Ur}}, \bibinfo {author} {\bibfnamefont {T.}~\bibnamefont
  {Mijatović}}, \bibinfo {author} {\bibfnamefont {D.~J.}\ \bibnamefont
  {Malenica}}, \bibinfo {author} {\bibfnamefont {N.}~\bibnamefont {Soić}}, \
  and\ \bibinfo {author} {\bibfnamefont {F.}~\bibnamefont {Haas}},\ }\bibinfo
  {title} {Multinucleon transfer reactions: Present status and perspectives},\
  \href {\doibase 10.1016/j.nimb.2013.04.093} {\bibfield  {journal} {\bibinfo
  {journal} {Nuclear Instruments and Methods in Physics Research Section B:
  Beam Interactions with Materials and Atoms}\ }\textbf {\bibinfo {volume}
  {317}},\ \bibinfo {pages} {743} (\bibinfo {year} {2013})}\BibitemShut
  {NoStop}%
\bibitem [{\citenamefont {Zhang}\ \emph {et~al.}(2018)\citenamefont {Zhang},
  \citenamefont {Li}, \citenamefont {Zhu},\ and\ \citenamefont
  {Wen}}]{Zhang2018}%
  \BibitemOpen
  \bibfield  {author} {\bibinfo {author} {\bibfnamefont {F.-S.}\ \bibnamefont
  {Zhang}}, \bibinfo {author} {\bibfnamefont {C.}~\bibnamefont {Li}}, \bibinfo
  {author} {\bibfnamefont {L.}~\bibnamefont {Zhu}}, \ and\ \bibinfo {author}
  {\bibfnamefont {P.}~\bibnamefont {Wen}},\ }\bibinfo {title} {Production cross
  sections for exotic nuclei with multinucleon transfer reactions},\ \href
  {\doibase 10.1007/s11467-018-0843-6} {\bibfield  {journal} {\bibinfo
  {journal} {Frontiers of Physics}\ }\textbf {\bibinfo {volume} {13}},\
  \bibinfo {pages} {132113} (\bibinfo {year} {2018})}\BibitemShut {NoStop}%
\bibitem [{\citenamefont {Adamian}\ \emph {et~al.}(2020)\citenamefont
  {Adamian}, \citenamefont {Antonenko}, \citenamefont {Diaz-Torres},\ and\
  \citenamefont {Heinz}}]{Adamian2020}%
  \BibitemOpen
  \bibfield  {author} {\bibinfo {author} {\bibfnamefont {G.~G.}\ \bibnamefont
  {Adamian}}, \bibinfo {author} {\bibfnamefont {N.~V.}\ \bibnamefont
  {Antonenko}}, \bibinfo {author} {\bibfnamefont {A.}~\bibnamefont
  {Diaz-Torres}}, \ and\ \bibinfo {author} {\bibfnamefont {S.}~\bibnamefont
  {Heinz}},\ }\bibinfo {title} {How to extend the chart of nuclides?},\ \href
  {\doibase 10.1140/epja/s10050-020-00046-7} {\bibfield  {journal} {\bibinfo
  {journal} {The European Physical Journal A}\ }\textbf {\bibinfo {volume}
  {56}},\ \bibinfo {pages} {47} (\bibinfo {year} {2020})}\BibitemShut {NoStop}%
\bibitem [{\citenamefont {Saiko}\ and\ \citenamefont
  {Karpov}(2022)}]{Saiko2022}%
  \BibitemOpen
  \bibfield  {author} {\bibinfo {author} {\bibfnamefont {V.}~\bibnamefont
  {Saiko}}\ and\ \bibinfo {author} {\bibfnamefont {A.}~\bibnamefont {Karpov}},\
  }\bibinfo {title} {Multinucleon transfer as a method for production of new
  heavy neutron-enriched isotopes of transuranium elements},\ \href {\doibase
  10.1140/epja/s10050-022-00688-9} {\bibfield  {journal} {\bibinfo  {journal}
  {The European Physical Journal A}\ }\textbf {\bibinfo {volume} {58}},\
  \bibinfo {pages} {41} (\bibinfo {year} {2022})}\BibitemShut {NoStop}%
\bibitem [{\citenamefont {Watanabe}\ \emph {et~al.}(2015)\citenamefont
  {Watanabe}, \citenamefont {Kim}, \citenamefont {Jeong}, \citenamefont
  {Hirayama}, \citenamefont {Imai}, \citenamefont {Ishiyama}, \citenamefont
  {Jung}, \citenamefont {Miyatake}, \citenamefont {Choi}, \citenamefont {Song},
  \citenamefont {Clement}, \citenamefont {de~France}, \citenamefont {Navin},
  \citenamefont {Rejmund}, \citenamefont {Schmitt}, \citenamefont {Pollarolo},
  \citenamefont {Corradi}, \citenamefont {Fioretto}, \citenamefont {Montanari},
  \citenamefont {Niikura}, \citenamefont {Suzuki}, \citenamefont {Nishibata},\
  and\ \citenamefont {Takatsu}}]{Watanabe2015}%
  \BibitemOpen
  \bibfield  {author} {\bibinfo {author} {\bibfnamefont {Y.~X.}\ \bibnamefont
  {Watanabe}}, \bibinfo {author} {\bibfnamefont {Y.~H.}\ \bibnamefont {Kim}},
  \bibinfo {author} {\bibfnamefont {S.~C.}\ \bibnamefont {Jeong}}, \bibinfo
  {author} {\bibfnamefont {Y.}~\bibnamefont {Hirayama}}, \bibinfo {author}
  {\bibfnamefont {N.}~\bibnamefont {Imai}}, \bibinfo {author} {\bibfnamefont
  {H.}~\bibnamefont {Ishiyama}}, \bibinfo {author} {\bibfnamefont {H.~S.}\
  \bibnamefont {Jung}}, \bibinfo {author} {\bibfnamefont {H.}~\bibnamefont
  {Miyatake}}, \bibinfo {author} {\bibfnamefont {S.}~\bibnamefont {Choi}},
  \bibinfo {author} {\bibfnamefont {J.~S.}\ \bibnamefont {Song}}, \bibinfo
  {author} {\bibfnamefont {E.}~\bibnamefont {Clement}}, \bibinfo {author}
  {\bibfnamefont {G.}~\bibnamefont {de~France}}, \bibinfo {author}
  {\bibfnamefont {A.}~\bibnamefont {Navin}}, \bibinfo {author} {\bibfnamefont
  {M.}~\bibnamefont {Rejmund}}, \bibinfo {author} {\bibfnamefont
  {C.}~\bibnamefont {Schmitt}}, \bibinfo {author} {\bibfnamefont
  {G.}~\bibnamefont {Pollarolo}}, \bibinfo {author} {\bibfnamefont
  {L.}~\bibnamefont {Corradi}}, \bibinfo {author} {\bibfnamefont
  {E.}~\bibnamefont {Fioretto}}, \bibinfo {author} {\bibfnamefont
  {D.}~\bibnamefont {Montanari}}, \bibinfo {author} {\bibfnamefont
  {M.}~\bibnamefont {Niikura}}, \bibinfo {author} {\bibfnamefont
  {D.}~\bibnamefont {Suzuki}}, \bibinfo {author} {\bibfnamefont
  {H.}~\bibnamefont {Nishibata}}, \ and\ \bibinfo {author} {\bibfnamefont
  {J.}~\bibnamefont {Takatsu}},\ }\bibinfo {title} {Pathway for the Production
  of Neutron-Rich Isotopes around the $N=126$ Shell Closure},\ \href {\doibase
  10.1103/PhysRevLett.115.172503} {\bibfield  {journal} {\bibinfo  {journal}
  {Phys. Rev. Lett.}\ }\textbf {\bibinfo {volume} {115}},\ \bibinfo {pages}
  {172503} (\bibinfo {year} {2015})}\BibitemShut {NoStop}%
\bibitem [{\citenamefont {Kozulin}\ \emph {et~al.}(2012)\citenamefont
  {Kozulin}, \citenamefont {Vardaci}, \citenamefont {Knyazheva}, \citenamefont
  {Bogachev}, \citenamefont {Dmitriev}, \citenamefont {Itkis}, \citenamefont
  {Itkis}, \citenamefont {Knyazev}, \citenamefont {Loktev}, \citenamefont
  {Novikov}, \citenamefont {Razinkov}, \citenamefont {Rudakov}, \citenamefont
  {Smirnov}, \citenamefont {Trzaska},\ and\ \citenamefont
  {Zagrebaev}}]{Kozulin2012}%
  \BibitemOpen
  \bibfield  {author} {\bibinfo {author} {\bibfnamefont {E.~M.}\ \bibnamefont
  {Kozulin}}, \bibinfo {author} {\bibfnamefont {E.}~\bibnamefont {Vardaci}},
  \bibinfo {author} {\bibfnamefont {G.~N.}\ \bibnamefont {Knyazheva}}, \bibinfo
  {author} {\bibfnamefont {A.~A.}\ \bibnamefont {Bogachev}}, \bibinfo {author}
  {\bibfnamefont {S.~N.}\ \bibnamefont {Dmitriev}}, \bibinfo {author}
  {\bibfnamefont {I.~M.}\ \bibnamefont {Itkis}}, \bibinfo {author}
  {\bibfnamefont {M.~G.}\ \bibnamefont {Itkis}}, \bibinfo {author}
  {\bibfnamefont {A.~G.}\ \bibnamefont {Knyazev}}, \bibinfo {author}
  {\bibfnamefont {T.~A.}\ \bibnamefont {Loktev}}, \bibinfo {author}
  {\bibfnamefont {K.~V.}\ \bibnamefont {Novikov}}, \bibinfo {author}
  {\bibfnamefont {E.~A.}\ \bibnamefont {Razinkov}}, \bibinfo {author}
  {\bibfnamefont {O.~V.}\ \bibnamefont {Rudakov}}, \bibinfo {author}
  {\bibfnamefont {S.~V.}\ \bibnamefont {Smirnov}}, \bibinfo {author}
  {\bibfnamefont {W.}~\bibnamefont {Trzaska}}, \ and\ \bibinfo {author}
  {\bibfnamefont {V.~I.}\ \bibnamefont {Zagrebaev}},\ }\bibinfo {title} {Mass
  distributions of the system ${}^{136}\mathrm{Xe}+{}^{208}$Pb at laboratory
  energies around the Coulomb barrier: A candidate reaction for the production
  of neutron-rich nuclei at $N=126$},\ \href {\doibase
  10.1103/PhysRevC.86.044611} {\bibfield  {journal} {\bibinfo  {journal} {Phys.
  Rev. C}\ }\textbf {\bibinfo {volume} {86}},\ \bibinfo {pages} {044611}
  (\bibinfo {year} {2012})}\BibitemShut {NoStop}%
\bibitem [{\citenamefont {Barrett}\ \emph {et~al.}(2015)\citenamefont
  {Barrett}, \citenamefont {Loveland}, \citenamefont {Yanez}, \citenamefont
  {Zhu}, \citenamefont {Ayangeakaa}, \citenamefont {Carpenter}, \citenamefont
  {Greene}, \citenamefont {Janssens}, \citenamefont {Lauritsen}, \citenamefont
  {McCutchan}, \citenamefont {Sonzogni}, \citenamefont {Chiara}, \citenamefont
  {Harker},\ and\ \citenamefont {Walters}}]{Barrett2015}%
  \BibitemOpen
  \bibfield  {author} {\bibinfo {author} {\bibfnamefont {J.~S.}\ \bibnamefont
  {Barrett}}, \bibinfo {author} {\bibfnamefont {W.}~\bibnamefont {Loveland}},
  \bibinfo {author} {\bibfnamefont {R.}~\bibnamefont {Yanez}}, \bibinfo
  {author} {\bibfnamefont {S.}~\bibnamefont {Zhu}}, \bibinfo {author}
  {\bibfnamefont {A.~D.}\ \bibnamefont {Ayangeakaa}}, \bibinfo {author}
  {\bibfnamefont {M.~P.}\ \bibnamefont {Carpenter}}, \bibinfo {author}
  {\bibfnamefont {J.~P.}\ \bibnamefont {Greene}}, \bibinfo {author}
  {\bibfnamefont {R.~V.~F.}\ \bibnamefont {Janssens}}, \bibinfo {author}
  {\bibfnamefont {T.}~\bibnamefont {Lauritsen}}, \bibinfo {author}
  {\bibfnamefont {E.~A.}\ \bibnamefont {McCutchan}}, \bibinfo {author}
  {\bibfnamefont {A.~A.}\ \bibnamefont {Sonzogni}}, \bibinfo {author}
  {\bibfnamefont {C.~J.}\ \bibnamefont {Chiara}}, \bibinfo {author}
  {\bibfnamefont {J.~L.}\ \bibnamefont {Harker}}, \ and\ \bibinfo {author}
  {\bibfnamefont {W.~B.}\ \bibnamefont {Walters}},\ }\bibinfo {title}
  {$^{136}\mathrm{Xe}+^{208}\mathrm{Pb}$ reaction: A test of models of
  multinucleon transfer reactions},\ \href {\doibase
  10.1103/PhysRevC.91.064615} {\bibfield  {journal} {\bibinfo  {journal} {Phys.
  Rev. C}\ }\textbf {\bibinfo {volume} {91}},\ \bibinfo {pages} {064615}
  (\bibinfo {year} {2015})}\BibitemShut {NoStop}%
\bibitem [{\citenamefont {Vogt}\ \emph {et~al.}(2015)\citenamefont {Vogt},
  \citenamefont {Birkenbach}, \citenamefont {Reiter}, \citenamefont {Corradi},
  \citenamefont {Mijatovi\ifmmode~\acute{c}\else \'{c}\fi{}}, \citenamefont
  {Montanari}, \citenamefont {Szilner}, \citenamefont {Bazzacco}, \citenamefont
  {Bowry}, \citenamefont {Bracco}, \citenamefont {Bruyneel}, \citenamefont
  {Crespi}, \citenamefont {de~Angelis}, \citenamefont {D\'esesquelles},
  \citenamefont {Eberth}, \citenamefont {Farnea}, \citenamefont {Fioretto},
  \citenamefont {Gadea}, \citenamefont {Geibel}, \citenamefont {Gengelbach},
  \citenamefont {Giaz}, \citenamefont {G\"orgen}, \citenamefont {Gottardo},
  \citenamefont {Grebosz}, \citenamefont {Hess}, \citenamefont {John},
  \citenamefont {Jolie}, \citenamefont {Judson}, \citenamefont {Jungclaus},
  \citenamefont {Korten}, \citenamefont {Leoni}, \citenamefont {Lunardi},
  \citenamefont {Menegazzo}, \citenamefont {Mengoni}, \citenamefont
  {Michelagnoli}, \citenamefont {Montagnoli}, \citenamefont {Napoli},
  \citenamefont {Pellegri}, \citenamefont {Pollarolo}, \citenamefont {Pullia},
  \citenamefont {Quintana}, \citenamefont {Radeck}, \citenamefont {Recchia},
  \citenamefont {Rosso}, \citenamefont {\ifmmode~\mbox{\c{S}}\else
  \c{S}\fi{}ahin}, \citenamefont {Salsac}, \citenamefont {Scarlassara},
  \citenamefont {S\"oderstr\"om}, \citenamefont {Stefanini}, \citenamefont
  {Steinbach}, \citenamefont {Stezowski}, \citenamefont {Szpak}, \citenamefont
  {Theisen}, \citenamefont {Ur}, \citenamefont {Valiente-Dob\'on},
  \citenamefont {Vandone},\ and\ \citenamefont {Wiens}}]{Vogt2015}%
  \BibitemOpen
  \bibfield  {author} {\bibinfo {author} {\bibfnamefont {A.}~\bibnamefont
  {Vogt}}, \bibinfo {author} {\bibfnamefont {B.}~\bibnamefont {Birkenbach}},
  \bibinfo {author} {\bibfnamefont {P.}~\bibnamefont {Reiter}}, \bibinfo
  {author} {\bibfnamefont {L.}~\bibnamefont {Corradi}}, \bibinfo {author}
  {\bibfnamefont {T.}~\bibnamefont {Mijatovi\ifmmode~\acute{c}\else
  \'{c}\fi{}}}, \bibinfo {author} {\bibfnamefont {D.}~\bibnamefont
  {Montanari}}, \bibinfo {author} {\bibfnamefont {S.}~\bibnamefont {Szilner}},
  \bibinfo {author} {\bibfnamefont {D.}~\bibnamefont {Bazzacco}}, \bibinfo
  {author} {\bibfnamefont {M.}~\bibnamefont {Bowry}}, \bibinfo {author}
  {\bibfnamefont {A.}~\bibnamefont {Bracco}}, \bibinfo {author} {\bibfnamefont
  {B.}~\bibnamefont {Bruyneel}}, \bibinfo {author} {\bibfnamefont {F.~C.~L.}\
  \bibnamefont {Crespi}}, \bibinfo {author} {\bibfnamefont {G.}~\bibnamefont
  {de~Angelis}}, \bibinfo {author} {\bibfnamefont {P.}~\bibnamefont
  {D\'esesquelles}}, \bibinfo {author} {\bibfnamefont {J.}~\bibnamefont
  {Eberth}}, \bibinfo {author} {\bibfnamefont {E.}~\bibnamefont {Farnea}},
  \bibinfo {author} {\bibfnamefont {E.}~\bibnamefont {Fioretto}}, \bibinfo
  {author} {\bibfnamefont {A.}~\bibnamefont {Gadea}}, \bibinfo {author}
  {\bibfnamefont {K.}~\bibnamefont {Geibel}}, \bibinfo {author} {\bibfnamefont
  {A.}~\bibnamefont {Gengelbach}}, \bibinfo {author} {\bibfnamefont
  {A.}~\bibnamefont {Giaz}}, \bibinfo {author} {\bibfnamefont {A.}~\bibnamefont
  {G\"orgen}}, \bibinfo {author} {\bibfnamefont {A.}~\bibnamefont {Gottardo}},
  \bibinfo {author} {\bibfnamefont {J.}~\bibnamefont {Grebosz}}, \bibinfo
  {author} {\bibfnamefont {H.}~\bibnamefont {Hess}}, \bibinfo {author}
  {\bibfnamefont {P.~R.}\ \bibnamefont {John}}, \bibinfo {author}
  {\bibfnamefont {J.}~\bibnamefont {Jolie}}, \bibinfo {author} {\bibfnamefont
  {D.~S.}\ \bibnamefont {Judson}}, \bibinfo {author} {\bibfnamefont
  {A.}~\bibnamefont {Jungclaus}}, \bibinfo {author} {\bibfnamefont
  {W.}~\bibnamefont {Korten}}, \bibinfo {author} {\bibfnamefont
  {S.}~\bibnamefont {Leoni}}, \bibinfo {author} {\bibfnamefont
  {S.}~\bibnamefont {Lunardi}}, \bibinfo {author} {\bibfnamefont
  {R.}~\bibnamefont {Menegazzo}}, \bibinfo {author} {\bibfnamefont
  {D.}~\bibnamefont {Mengoni}}, \bibinfo {author} {\bibfnamefont
  {C.}~\bibnamefont {Michelagnoli}}, \bibinfo {author} {\bibfnamefont
  {G.}~\bibnamefont {Montagnoli}}, \bibinfo {author} {\bibfnamefont
  {D.}~\bibnamefont {Napoli}}, \bibinfo {author} {\bibfnamefont
  {L.}~\bibnamefont {Pellegri}}, \bibinfo {author} {\bibfnamefont
  {G.}~\bibnamefont {Pollarolo}}, \bibinfo {author} {\bibfnamefont
  {A.}~\bibnamefont {Pullia}}, \bibinfo {author} {\bibfnamefont
  {B.}~\bibnamefont {Quintana}}, \bibinfo {author} {\bibfnamefont
  {F.}~\bibnamefont {Radeck}}, \bibinfo {author} {\bibfnamefont
  {F.}~\bibnamefont {Recchia}}, \bibinfo {author} {\bibfnamefont
  {D.}~\bibnamefont {Rosso}}, \bibinfo {author} {\bibfnamefont
  {E.}~\bibnamefont {\ifmmode~\mbox{\c{S}}\else \c{S}\fi{}ahin}}, \bibinfo
  {author} {\bibfnamefont {M.~D.}\ \bibnamefont {Salsac}}, \bibinfo {author}
  {\bibfnamefont {F.}~\bibnamefont {Scarlassara}}, \bibinfo {author}
  {\bibfnamefont {P.-A.}\ \bibnamefont {S\"oderstr\"om}}, \bibinfo {author}
  {\bibfnamefont {A.~M.}\ \bibnamefont {Stefanini}}, \bibinfo {author}
  {\bibfnamefont {T.}~\bibnamefont {Steinbach}}, \bibinfo {author}
  {\bibfnamefont {O.}~\bibnamefont {Stezowski}}, \bibinfo {author}
  {\bibfnamefont {B.}~\bibnamefont {Szpak}}, \bibinfo {author} {\bibfnamefont
  {C.}~\bibnamefont {Theisen}}, \bibinfo {author} {\bibfnamefont
  {C.}~\bibnamefont {Ur}}, \bibinfo {author} {\bibfnamefont {J.~J.}\
  \bibnamefont {Valiente-Dob\'on}}, \bibinfo {author} {\bibfnamefont
  {V.}~\bibnamefont {Vandone}}, \ and\ \bibinfo {author} {\bibfnamefont
  {A.}~\bibnamefont {Wiens}},\ }\bibinfo {title} {Light and heavy transfer
  products in $^{136}\mathrm{Xe}+^{238}\mathrm{U}$ multinucleon transfer
  reactions},\ \href {\doibase 10.1103/PhysRevC.92.024619} {\bibfield
  {journal} {\bibinfo  {journal} {Phys. Rev. C}\ }\textbf {\bibinfo {volume}
  {92}},\ \bibinfo {pages} {024619} (\bibinfo {year} {2015})}\BibitemShut
  {NoStop}%
\bibitem [{\citenamefont {Kozulin}\ \emph {et~al.}(2017)\citenamefont
  {Kozulin}, \citenamefont {Zagrebaev}, \citenamefont {Knyazheva},
  \citenamefont {Itkis}, \citenamefont {Novikov}, \citenamefont {Itkis},
  \citenamefont {Dmitriev}, \citenamefont {Harca}, \citenamefont
  {Bondarchenko}, \citenamefont {Karpov}, \citenamefont {Saiko},\ and\
  \citenamefont {Vardaci}}]{Kozulin2017}%
  \BibitemOpen
  \bibfield  {author} {\bibinfo {author} {\bibfnamefont {E.~M.}\ \bibnamefont
  {Kozulin}}, \bibinfo {author} {\bibfnamefont {V.~I.}\ \bibnamefont
  {Zagrebaev}}, \bibinfo {author} {\bibfnamefont {G.~N.}\ \bibnamefont
  {Knyazheva}}, \bibinfo {author} {\bibfnamefont {I.~M.}\ \bibnamefont
  {Itkis}}, \bibinfo {author} {\bibfnamefont {K.~V.}\ \bibnamefont {Novikov}},
  \bibinfo {author} {\bibfnamefont {M.~G.}\ \bibnamefont {Itkis}}, \bibinfo
  {author} {\bibfnamefont {S.~N.}\ \bibnamefont {Dmitriev}}, \bibinfo {author}
  {\bibfnamefont {I.~M.}\ \bibnamefont {Harca}}, \bibinfo {author}
  {\bibfnamefont {A.~E.}\ \bibnamefont {Bondarchenko}}, \bibinfo {author}
  {\bibfnamefont {A.~V.}\ \bibnamefont {Karpov}}, \bibinfo {author}
  {\bibfnamefont {V.~V.}\ \bibnamefont {Saiko}}, \ and\ \bibinfo {author}
  {\bibfnamefont {E.}~\bibnamefont {Vardaci}},\ }\bibinfo {title} {Inverse
  quasifission in the reactions $^{156,160}\mathrm{Gd}+^{186}\mathrm{W}$},\
  \href {\doibase 10.1103/PhysRevC.96.064621} {\bibfield  {journal} {\bibinfo
  {journal} {Phys. Rev. C}\ }\textbf {\bibinfo {volume} {96}},\ \bibinfo
  {pages} {064621} (\bibinfo {year} {2017})}\BibitemShut {NoStop}%
\bibitem [{\citenamefont {Dikli\ifmmode~\acute{c}\else \'{c}\fi{}}\ \emph
  {et~al.}(2023)\citenamefont {Dikli\ifmmode~\acute{c}\else \'{c}\fi{}},
  \citenamefont {Szilner}, \citenamefont {Corradi}, \citenamefont
  {Mijatovi\ifmmode~\acute{c}\else \'{c}\fi{}}, \citenamefont {Pollarolo},
  \citenamefont {\ifmmode \check{C}\else \v{C}\fi{}olovi\ifmmode~\acute{c}\else
  \'{c}\fi{}}, \citenamefont {Colucci}, \citenamefont {Fioretto}, \citenamefont
  {Galtarossa}, \citenamefont {Goasduff}, \citenamefont {Gottardo},
  \citenamefont {Grebosz}, \citenamefont {Illana}, \citenamefont {Jaworski},
  \citenamefont {Gomez}, \citenamefont {Marchi}, \citenamefont {Mengoni},
  \citenamefont {Montagnoli}, \citenamefont {Nurki\ifmmode~\acute{c}\else
  \'{c}\fi{}}, \citenamefont {Siciliano}, \citenamefont
  {Soi\ifmmode~\acute{c}\else \'{c}\fi{}}, \citenamefont {Stefanini},
  \citenamefont {Testov}, \citenamefont {Valiente-Dob\'on},\ and\ \citenamefont
  {Vukman}}]{Diklic2023}%
  \BibitemOpen
  \bibfield  {author} {\bibinfo {author} {\bibfnamefont {J.}~\bibnamefont
  {Dikli\ifmmode~\acute{c}\else \'{c}\fi{}}}, \bibinfo {author} {\bibfnamefont
  {S.}~\bibnamefont {Szilner}}, \bibinfo {author} {\bibfnamefont
  {L.}~\bibnamefont {Corradi}}, \bibinfo {author} {\bibfnamefont
  {T.}~\bibnamefont {Mijatovi\ifmmode~\acute{c}\else \'{c}\fi{}}}, \bibinfo
  {author} {\bibfnamefont {G.}~\bibnamefont {Pollarolo}}, \bibinfo {author}
  {\bibfnamefont {P.}~\bibnamefont {\ifmmode \check{C}\else
  \v{C}\fi{}olovi\ifmmode~\acute{c}\else \'{c}\fi{}}}, \bibinfo {author}
  {\bibfnamefont {G.}~\bibnamefont {Colucci}}, \bibinfo {author} {\bibfnamefont
  {E.}~\bibnamefont {Fioretto}}, \bibinfo {author} {\bibfnamefont
  {F.}~\bibnamefont {Galtarossa}}, \bibinfo {author} {\bibfnamefont
  {A.}~\bibnamefont {Goasduff}}, \bibinfo {author} {\bibfnamefont
  {A.}~\bibnamefont {Gottardo}}, \bibinfo {author} {\bibfnamefont
  {J.}~\bibnamefont {Grebosz}}, \bibinfo {author} {\bibfnamefont
  {A.}~\bibnamefont {Illana}}, \bibinfo {author} {\bibfnamefont
  {G.}~\bibnamefont {Jaworski}}, \bibinfo {author} {\bibfnamefont {M.~J.}\
  \bibnamefont {Gomez}}, \bibinfo {author} {\bibfnamefont {T.}~\bibnamefont
  {Marchi}}, \bibinfo {author} {\bibfnamefont {D.}~\bibnamefont {Mengoni}},
  \bibinfo {author} {\bibfnamefont {G.}~\bibnamefont {Montagnoli}}, \bibinfo
  {author} {\bibfnamefont {D.}~\bibnamefont {Nurki\ifmmode~\acute{c}\else
  \'{c}\fi{}}}, \bibinfo {author} {\bibfnamefont {M.}~\bibnamefont
  {Siciliano}}, \bibinfo {author} {\bibfnamefont {N.}~\bibnamefont
  {Soi\ifmmode~\acute{c}\else \'{c}\fi{}}}, \bibinfo {author} {\bibfnamefont
  {A.~M.}\ \bibnamefont {Stefanini}}, \bibinfo {author} {\bibfnamefont
  {D.}~\bibnamefont {Testov}}, \bibinfo {author} {\bibfnamefont {J.~J.}\
  \bibnamefont {Valiente-Dob\'on}}, \ and\ \bibinfo {author} {\bibfnamefont
  {N.}~\bibnamefont {Vukman}},\ }\bibinfo {title} {Transfer reactions in
  $^{206}\mathrm{Pb}+^{118}\mathrm{Sn}$: From quasielastic to deep-inelastic
  processes},\ \href {\doibase 10.1103/PhysRevC.107.014619} {\bibfield
  {journal} {\bibinfo  {journal} {Phys. Rev. C}\ }\textbf {\bibinfo {volume}
  {107}},\ \bibinfo {pages} {014619} (\bibinfo {year} {2023})}\BibitemShut
  {NoStop}%
\bibitem [{\citenamefont {Wilczyński}(1973)}]{Wilczy1973}%
  \BibitemOpen
  \bibfield  {author} {\bibinfo {author} {\bibfnamefont {J.}~\bibnamefont
  {Wilczyński}},\ }\bibinfo {title} {Nuclear molecules and nuclear friction},\
  \href {\doibase 10.1016/0370-2693(73)90021-X} {\bibfield  {journal} {\bibinfo
   {journal} {Physics Letters B}\ }\textbf {\bibinfo {volume} {47}},\ \bibinfo
  {pages} {484} (\bibinfo {year} {1973})}\BibitemShut {NoStop}%
\bibitem [{\citenamefont {Heinz}\ and\ \citenamefont
  {Devaraja}(2022)}]{Heinz2022}%
  \BibitemOpen
  \bibfield  {author} {\bibinfo {author} {\bibfnamefont {S.}~\bibnamefont
  {Heinz}}\ and\ \bibinfo {author} {\bibfnamefont {H.~M.}\ \bibnamefont
  {Devaraja}},\ }\bibinfo {title} {Nucleosynthesis in multinucleon transfer
  reactions},\ \href {\doibase 10.1140/epja/s10050-022-00771-1} {\bibfield
  {journal} {\bibinfo  {journal} {The European Physical Journal A}\ }\textbf
  {\bibinfo {volume} {58}},\ \bibinfo {pages} {114} (\bibinfo {year}
  {2022})}\BibitemShut {NoStop}%
\bibitem [{\citenamefont {Valverde}\ \emph {et~al.}(2020)\citenamefont
  {Valverde}, \citenamefont {Brodeur}, \citenamefont {Clark}, \citenamefont
  {Lascar},\ and\ \citenamefont {Savard}}]{Valverde2020}%
  \BibitemOpen
  \bibfield  {author} {\bibinfo {author} {\bibfnamefont {A.}~\bibnamefont
  {Valverde}}, \bibinfo {author} {\bibfnamefont {M.}~\bibnamefont {Brodeur}},
  \bibinfo {author} {\bibfnamefont {J.}~\bibnamefont {Clark}}, \bibinfo
  {author} {\bibfnamefont {D.}~\bibnamefont {Lascar}}, \ and\ \bibinfo {author}
  {\bibfnamefont {G.}~\bibnamefont {Savard}},\ }\bibinfo {title} {A
  cooler-buncher for the N=126 factory at Argonne National Laboratory},\ \href
  {\doibase https://doi.org/10.1016/j.nimb.2019.04.070} {\bibfield  {journal}
  {\bibinfo  {journal} {Nuclear Instruments and Methods in Physics Research
  Section B: Beam Interactions with Materials and Atoms}\ }\textbf {\bibinfo
  {volume} {463}},\ \bibinfo {pages} {330} (\bibinfo {year}
  {2020})}\BibitemShut {NoStop}%
\bibitem [{\citenamefont {Karpov}\ and\ \citenamefont
  {Saiko}(2017)}]{Karpov2017}%
  \BibitemOpen
  \bibfield  {author} {\bibinfo {author} {\bibfnamefont {A.~V.}\ \bibnamefont
  {Karpov}}\ and\ \bibinfo {author} {\bibfnamefont {V.~V.}\ \bibnamefont
  {Saiko}},\ }\bibinfo {title} {Modeling near-barrier collisions of heavy ions
  based on a Langevin-type approach},\ \href {\doibase
  10.1103/PhysRevC.96.024618} {\bibfield  {journal} {\bibinfo  {journal}
  {Physical Review C}\ }\textbf {\bibinfo {volume} {96}},\ \bibinfo {pages}
  {024618} (\bibinfo {year} {2017})}\BibitemShut {NoStop}%
\bibitem [{\citenamefont {Saiko}\ and\ \citenamefont
  {Karpov}(2019)}]{Saiko2019}%
  \BibitemOpen
  \bibfield  {author} {\bibinfo {author} {\bibfnamefont {V.~V.}\ \bibnamefont
  {Saiko}}\ and\ \bibinfo {author} {\bibfnamefont {A.~V.}\ \bibnamefont
  {Karpov}},\ }\bibinfo {title} {Analysis of multinucleon transfer reactions
  with spherical and statically deformed nuclei using a Langevin-type
  approach},\ \href {\doibase 10.1103/PhysRevC.99.014613} {\bibfield  {journal}
  {\bibinfo  {journal} {Physical Review C}\ }\textbf {\bibinfo {volume} {99}},\
  \bibinfo {pages} {014613} (\bibinfo {year} {2019})}\BibitemShut {NoStop}%
\bibitem [{\citenamefont {Feng}(2017)}]{Feng2017}%
  \BibitemOpen
  \bibfield  {author} {\bibinfo {author} {\bibfnamefont {Z.-Q.}\ \bibnamefont
  {Feng}},\ }\bibinfo {title} {Production of neutron-rich isotopes around
  $N=126$ in multinucleon transfer reactions},\ \href {\doibase
  10.1103/PhysRevC.95.024615} {\bibfield  {journal} {\bibinfo  {journal} {Phys.
  Rev. C}\ }\textbf {\bibinfo {volume} {95}},\ \bibinfo {pages} {024615}
  (\bibinfo {year} {2017})}\BibitemShut {NoStop}%
\bibitem [{\citenamefont {Bao}\ \emph {et~al.}(2018)\citenamefont {Bao},
  \citenamefont {Guo}, \citenamefont {Li},\ and\ \citenamefont
  {Zhang}}]{Bao2018}%
  \BibitemOpen
  \bibfield  {author} {\bibinfo {author} {\bibfnamefont {X.~J.}\ \bibnamefont
  {Bao}}, \bibinfo {author} {\bibfnamefont {S.~Q.}\ \bibnamefont {Guo}},
  \bibinfo {author} {\bibfnamefont {J.~Q.}\ \bibnamefont {Li}}, \ and\ \bibinfo
  {author} {\bibfnamefont {H.~F.}\ \bibnamefont {Zhang}},\ }\bibinfo {title}
  {Influence of neutron excess of projectile on multinucleon transfer
  reactions},\ \href {\doibase 10.1016/j.physletb.2018.08.049} {\bibfield
  {journal} {\bibinfo  {journal} {Physics Letters B}\ }\textbf {\bibinfo
  {volume} {785}},\ \bibinfo {pages} {221} (\bibinfo {year}
  {2018})}\BibitemShut {NoStop}%
\bibitem [{\citenamefont {Zhu}\ \emph {et~al.}(2018)\citenamefont {Zhu},
  \citenamefont {Wen}, \citenamefont {Lin}, \citenamefont {Bao}, \citenamefont
  {Su}, \citenamefont {Li},\ and\ \citenamefont {Guo}}]{Zhu2018}%
  \BibitemOpen
  \bibfield  {author} {\bibinfo {author} {\bibfnamefont {L.}~\bibnamefont
  {Zhu}}, \bibinfo {author} {\bibfnamefont {P.-W.}\ \bibnamefont {Wen}},
  \bibinfo {author} {\bibfnamefont {C.-J.}\ \bibnamefont {Lin}}, \bibinfo
  {author} {\bibfnamefont {X.-J.}\ \bibnamefont {Bao}}, \bibinfo {author}
  {\bibfnamefont {J.}~\bibnamefont {Su}}, \bibinfo {author} {\bibfnamefont
  {C.}~\bibnamefont {Li}}, \ and\ \bibinfo {author} {\bibfnamefont {C.-C.}\
  \bibnamefont {Guo}},\ }\bibinfo {title} {Shell effects in a multinucleon
  transfer process},\ \href {\doibase 10.1103/PhysRevC.97.044614} {\bibfield
  {journal} {\bibinfo  {journal} {Physical Review C}\ }\textbf {\bibinfo
  {volume} {97}},\ \bibinfo {pages} {044614} (\bibinfo {year}
  {2018})}\BibitemShut {NoStop}%
\bibitem [{\citenamefont {Guo}\ \emph {et~al.}(2019)\citenamefont {Guo},
  \citenamefont {Bao}, \citenamefont {Zhang}, \citenamefont {Li},\ and\
  \citenamefont {Wang}}]{Guo2019}%
  \BibitemOpen
  \bibfield  {author} {\bibinfo {author} {\bibfnamefont {S.~Q.}\ \bibnamefont
  {Guo}}, \bibinfo {author} {\bibfnamefont {X.~J.}\ \bibnamefont {Bao}},
  \bibinfo {author} {\bibfnamefont {H.~F.}\ \bibnamefont {Zhang}}, \bibinfo
  {author} {\bibfnamefont {J.~Q.}\ \bibnamefont {Li}}, \ and\ \bibinfo {author}
  {\bibfnamefont {N.}~\bibnamefont {Wang}},\ }\bibinfo {title} {Effect of
  dynamical deformation on the production distribution in multinucleon transfer
  reactions},\ \href {\doibase 10.1103/PhysRevC.100.054616} {\bibfield
  {journal} {\bibinfo  {journal} {Physical Review C}\ }\textbf {\bibinfo
  {volume} {100}},\ \bibinfo {pages} {054616} (\bibinfo {year}
  {2019})}\BibitemShut {NoStop}%
\bibitem [{\citenamefont {Li}\ \emph {et~al.}(2019)\citenamefont {Li},
  \citenamefont {Xu}, \citenamefont {Li}, \citenamefont {Zhang}, \citenamefont
  {Li}, \citenamefont {Sokhna}, \citenamefont {Ge}, \citenamefont {Zhang},
  \citenamefont {Wen},\ and\ \citenamefont {Zhang}}]{Li2019}%
  \BibitemOpen
  \bibfield  {author} {\bibinfo {author} {\bibfnamefont {C.}~\bibnamefont
  {Li}}, \bibinfo {author} {\bibfnamefont {X.}~\bibnamefont {Xu}}, \bibinfo
  {author} {\bibfnamefont {J.}~\bibnamefont {Li}}, \bibinfo {author}
  {\bibfnamefont {G.}~\bibnamefont {Zhang}}, \bibinfo {author} {\bibfnamefont
  {B.}~\bibnamefont {Li}}, \bibinfo {author} {\bibfnamefont {C.~A.~T.}\
  \bibnamefont {Sokhna}}, \bibinfo {author} {\bibfnamefont {Z.}~\bibnamefont
  {Ge}}, \bibinfo {author} {\bibfnamefont {F.}~\bibnamefont {Zhang}}, \bibinfo
  {author} {\bibfnamefont {P.}~\bibnamefont {Wen}}, \ and\ \bibinfo {author}
  {\bibfnamefont {F.-S.}\ \bibnamefont {Zhang}},\ }\bibinfo {title} {Production
  of new neutron-rich heavy nuclei with $Z=56-80$ in the multinucleon transfer
  reactions of $^{136}\mathrm{Xe}+^{198}\mathrm{Pt}$},\ \href {\doibase
  10.1103/PhysRevC.99.024602} {\bibfield  {journal} {\bibinfo  {journal} {Phys.
  Rev. C}\ }\textbf {\bibinfo {volume} {99}},\ \bibinfo {pages} {024602}
  (\bibinfo {year} {2019})}\BibitemShut {NoStop}%
\bibitem [{\citenamefont {Zhao}\ \emph {et~al.}(2021)\citenamefont {Zhao},
  \citenamefont {Liu}, \citenamefont {Zhang},\ and\ \citenamefont
  {Wang}}]{Zhao2021}%
  \BibitemOpen
  \bibfield  {author} {\bibinfo {author} {\bibfnamefont {K.}~\bibnamefont
  {Zhao}}, \bibinfo {author} {\bibfnamefont {Z.}~\bibnamefont {Liu}}, \bibinfo
  {author} {\bibfnamefont {F.}~\bibnamefont {Zhang}}, \ and\ \bibinfo {author}
  {\bibfnamefont {N.}~\bibnamefont {Wang}},\ }\bibinfo {title} {Production of
  neutron-rich N=126 nuclei in multinucleon transfer reactions: Comparison
  between $^{136}\mathrm{Xe}+^{198}\mathrm{Pt}$ and
  $^{238}\mathrm{U}+^{198}\mathrm{Pt}$ reactions},\ \href {\doibase
  10.1016/j.physletb.2021.136101} {\bibfield  {journal} {\bibinfo  {journal}
  {Physics Letters B}\ }\textbf {\bibinfo {volume} {815}},\ \bibinfo {pages}
  {136101} (\bibinfo {year} {2021})}\BibitemShut {NoStop}%
\bibitem [{\citenamefont {Zhao}\ \emph {et~al.}(2022)\citenamefont {Zhao},
  \citenamefont {Liu}, \citenamefont {Zhang}, \citenamefont {Wang},\ and\
  \citenamefont {Duan}}]{Zhao2022}%
  \BibitemOpen
  \bibfield  {author} {\bibinfo {author} {\bibfnamefont {K.}~\bibnamefont
  {Zhao}}, \bibinfo {author} {\bibfnamefont {Z.}~\bibnamefont {Liu}}, \bibinfo
  {author} {\bibfnamefont {F.~S.}\ \bibnamefont {Zhang}}, \bibinfo {author}
  {\bibfnamefont {N.}~\bibnamefont {Wang}}, \ and\ \bibinfo {author}
  {\bibfnamefont {J.~Z.}\ \bibnamefont {Duan}},\ }\bibinfo {title} {Distinct
  sequential and massive transfer processes for production of neutron-rich
  $N\ensuremath{\approx}126$ nuclei in $^{238}\mathrm{U}+^{198}\mathrm{Pt}$
  reactions},\ \href {\doibase 10.1103/PhysRevC.106.L011602} {\bibfield
  {journal} {\bibinfo  {journal} {Phys. Rev. C}\ }\textbf {\bibinfo {volume}
  {106}},\ \bibinfo {pages} {L011602} (\bibinfo {year} {2022})}\BibitemShut
  {NoStop}%
\bibitem [{\citenamefont {Sekizawa}\ and\ \citenamefont
  {Yabana}(2016)}]{Sekizawa2016}%
  \BibitemOpen
  \bibfield  {author} {\bibinfo {author} {\bibfnamefont {K.}~\bibnamefont
  {Sekizawa}}\ and\ \bibinfo {author} {\bibfnamefont {K.}~\bibnamefont
  {Yabana}},\ }\bibinfo {title} {Time-dependent Hartree-Fock calculations for
  multinucleon transfer and quasifission processes in the
  $^{64}\text{Ni}+^{238}\text{U}$ reaction},\ \href {\doibase
  10.1103/PhysRevC.93.054616} {\bibfield  {journal} {\bibinfo  {journal} {Phys.
  Rev. C}\ }\textbf {\bibinfo {volume} {93}},\ \bibinfo {pages} {054616}
  (\bibinfo {year} {2016})}\BibitemShut {NoStop}%
\bibitem [{\citenamefont {Jiang}\ and\ \citenamefont {Wang}(2018)}]{Jiang2018}%
  \BibitemOpen
  \bibfield  {author} {\bibinfo {author} {\bibfnamefont {X.}~\bibnamefont
  {Jiang}}\ and\ \bibinfo {author} {\bibfnamefont {N.}~\bibnamefont {Wang}},\
  }\bibinfo {title} {Production mechanism of neutron-rich nuclei around N = 126
  in the multi-nucleon transfer reaction
  $^{132}\mathrm{Sn}+^{208}\mathrm{Pb*}$},\ \href {\doibase
  10.1088/1674-1137/42/10/104105} {\bibfield  {journal} {\bibinfo  {journal}
  {Chinese Physics C}\ }\textbf {\bibinfo {volume} {42}},\ \bibinfo {pages}
  {104105} (\bibinfo {year} {2018})}\BibitemShut {NoStop}%
\bibitem [{\citenamefont {Sun}\ and\ \citenamefont {Guo}(2023)}]{Sun2023}%
  \BibitemOpen
  \bibfield  {author} {\bibinfo {author} {\bibfnamefont {X.-X.}\ \bibnamefont
  {Sun}}\ and\ \bibinfo {author} {\bibfnamefont {L.}~\bibnamefont {Guo}},\
  }\bibinfo {title} {Microscopic study of fusion reactions with a weakly bound
  nucleus: Effects of deformed halo},\ \href {\doibase
  10.1103/PhysRevC.107.L011601} {\bibfield  {journal} {\bibinfo  {journal}
  {Physical Review C}\ }\textbf {\bibinfo {volume} {107}},\ \bibinfo {pages}
  {L011601} (\bibinfo {year} {2023})}\BibitemShut {NoStop}%
\bibitem [{\citenamefont {Ayik}\ \emph {et~al.}(2021)\citenamefont {Ayik},
  \citenamefont {Arik}, \citenamefont {Karanfil}, \citenamefont {Yilmaz},
  \citenamefont {Yilmaz},\ and\ \citenamefont {Umar}}]{Ayik2021}%
  \BibitemOpen
  \bibfield  {author} {\bibinfo {author} {\bibfnamefont {S.}~\bibnamefont
  {Ayik}}, \bibinfo {author} {\bibfnamefont {M.}~\bibnamefont {Arik}}, \bibinfo
  {author} {\bibfnamefont {E.~C.}\ \bibnamefont {Karanfil}}, \bibinfo {author}
  {\bibfnamefont {O.}~\bibnamefont {Yilmaz}}, \bibinfo {author} {\bibfnamefont
  {B.}~\bibnamefont {Yilmaz}}, \ and\ \bibinfo {author} {\bibfnamefont {A.~S.}\
  \bibnamefont {Umar}},\ }\bibinfo {title} {Quantal diffusion description of
  isotope production via the multinucleon transfer mechanism in
  $^{48}\mathrm{Ca}+^{238}\mathrm{U}$ collisions},\ \href {\doibase
  10.1103/PhysRevC.104.054614} {\bibfield  {journal} {\bibinfo  {journal}
  {Phys. Rev. C}\ }\textbf {\bibinfo {volume} {104}},\ \bibinfo {pages}
  {054614} (\bibinfo {year} {2021})}\BibitemShut {NoStop}%
\bibitem [{\citenamefont {Ayik}\ \emph {et~al.}(2023)\citenamefont {Ayik},
  \citenamefont {Arik}, \citenamefont {Yilmaz}, \citenamefont {Yilmaz},\ and\
  \citenamefont {Umar}}]{Ayik2023}%
  \BibitemOpen
  \bibfield  {author} {\bibinfo {author} {\bibfnamefont {S.}~\bibnamefont
  {Ayik}}, \bibinfo {author} {\bibfnamefont {M.}~\bibnamefont {Arik}}, \bibinfo
  {author} {\bibfnamefont {O.}~\bibnamefont {Yilmaz}}, \bibinfo {author}
  {\bibfnamefont {B.}~\bibnamefont {Yilmaz}}, \ and\ \bibinfo {author}
  {\bibfnamefont {A.~S.}\ \bibnamefont {Umar}},\ }\bibinfo {title}
  {Multinucleon transfer mechanism in $^{250}\mathrm{Cf}+^{232}\mathrm{Th}$
  collisions using the quantal transport description based on the stochastic
  mean-field approach},\ \href {\doibase 10.1103/PhysRevC.107.014609}
  {\bibfield  {journal} {\bibinfo  {journal} {Phys. Rev. C}\ }\textbf {\bibinfo
  {volume} {107}},\ \bibinfo {pages} {014609} (\bibinfo {year}
  {2023})}\BibitemShut {NoStop}%
\bibitem [{\citenamefont {Adamian}\ \emph {et~al.}(1997)\citenamefont
  {Adamian}, \citenamefont {Antonenko}, \citenamefont {Scheid},\ and\
  \citenamefont {Volkov}}]{Adamian1997}%
  \BibitemOpen
  \bibfield  {author} {\bibinfo {author} {\bibfnamefont {G.}~\bibnamefont
  {Adamian}}, \bibinfo {author} {\bibfnamefont {N.}~\bibnamefont {Antonenko}},
  \bibinfo {author} {\bibfnamefont {W.}~\bibnamefont {Scheid}}, \ and\ \bibinfo
  {author} {\bibfnamefont {V.}~\bibnamefont {Volkov}},\ }\bibinfo {title}
  {Treatment of competition between complete fusion and quasifission in
  collisions of heavy nuclei},\ \href {\doibase 10.1016/S0375-9474(97)00605-2}
  {\bibfield  {journal} {\bibinfo  {journal} {Nuclear Physics A}\ }\textbf
  {\bibinfo {volume} {627}},\ \bibinfo {pages} {361} (\bibinfo {year}
  {1997})}\BibitemShut {NoStop}%
\bibitem [{\citenamefont {Li}\ \emph {et~al.}(2003)\citenamefont {Li},
  \citenamefont {Wang}, \citenamefont {Li}, \citenamefont {Xu}, \citenamefont
  {Zuo}, \citenamefont {Zhao}, \citenamefont {Li},\ and\ \citenamefont
  {Scheid}}]{Li2003}%
  \BibitemOpen
  \bibfield  {author} {\bibinfo {author} {\bibfnamefont {W.}~\bibnamefont
  {Li}}, \bibinfo {author} {\bibfnamefont {N.}~\bibnamefont {Wang}}, \bibinfo
  {author} {\bibfnamefont {J.~F.}\ \bibnamefont {Li}}, \bibinfo {author}
  {\bibfnamefont {H.}~\bibnamefont {Xu}}, \bibinfo {author} {\bibfnamefont
  {W.}~\bibnamefont {Zuo}}, \bibinfo {author} {\bibfnamefont {E.}~\bibnamefont
  {Zhao}}, \bibinfo {author} {\bibfnamefont {J.~Q.}\ \bibnamefont {Li}}, \ and\
  \bibinfo {author} {\bibfnamefont {W.}~\bibnamefont {Scheid}},\ }\bibinfo
  {title} {Fusion probability in heavy-ion collisions by a dinuclear-system
  model},\ \href {\doibase 10.1209/epl/i2003-00622-0} {\bibfield  {journal}
  {\bibinfo  {journal} {Europhysics Letters (EPL)}\ }\textbf {\bibinfo {volume}
  {64}},\ \bibinfo {pages} {750} (\bibinfo {year} {2003})}\BibitemShut
  {NoStop}%
\bibitem [{\citenamefont {Feng}\ \emph {et~al.}(2007)\citenamefont {Feng},
  \citenamefont {Jin}, \citenamefont {Li},\ and\ \citenamefont
  {Scheid}}]{Feng2007}%
  \BibitemOpen
  \bibfield  {author} {\bibinfo {author} {\bibfnamefont {Z.-Q.}\ \bibnamefont
  {Feng}}, \bibinfo {author} {\bibfnamefont {G.-M.}\ \bibnamefont {Jin}},
  \bibinfo {author} {\bibfnamefont {J.-Q.}\ \bibnamefont {Li}}, \ and\ \bibinfo
  {author} {\bibfnamefont {W.}~\bibnamefont {Scheid}},\ }\bibinfo {title}
  {Formation of superheavy nuclei in cold fusion reactions},\ \href {\doibase
  10.1103/PhysRevC.76.044606} {\bibfield  {journal} {\bibinfo  {journal}
  {Physical Review C}\ }\textbf {\bibinfo {volume} {76}},\ \bibinfo {pages}
  {044606} (\bibinfo {year} {2007})}\BibitemShut {NoStop}%
\bibitem [{\citenamefont {Zhu}\ and\ \citenamefont
  {Su}(2021)}]{Zhu2021unified}%
  \BibitemOpen
  \bibfield  {author} {\bibinfo {author} {\bibfnamefont {L.}~\bibnamefont
  {Zhu}}\ and\ \bibinfo {author} {\bibfnamefont {J.}~\bibnamefont {Su}},\
  }\bibinfo {title} {Unified description of fusion and multinucleon transfer
  processes within the dinuclear system model},\ \href {\doibase
  10.1103/PhysRevC.104.044606} {\bibfield  {journal} {\bibinfo  {journal}
  {Physical Review C}\ }\textbf {\bibinfo {volume} {104}},\ \bibinfo {pages}
  {044606} (\bibinfo {year} {2021})}\BibitemShut {NoStop}%
\bibitem [{\citenamefont {Zhu}(2021)}]{Zhu2021}%
  \BibitemOpen
  \bibfield  {author} {\bibinfo {author} {\bibfnamefont {L.}~\bibnamefont
  {Zhu}},\ }\bibinfo {title} {Shell inhibition on production of N=126 isotones
  in multinucleon transfer reactions},\ \href {\doibase
  10.1016/j.physletb.2021.136226} {\bibfield  {journal} {\bibinfo  {journal}
  {Physics Letters B}\ }\textbf {\bibinfo {volume} {816}},\ \bibinfo {pages}
  {136226} (\bibinfo {year} {2021})}\BibitemShut {NoStop}%
\bibitem [{\citenamefont {Zhu}\ \emph {et~al.}(2022)\citenamefont {Zhu},
  \citenamefont {Su}, \citenamefont {Li},\ and\ \citenamefont
  {Zhang}}]{Zhu2022}%
  \BibitemOpen
  \bibfield  {author} {\bibinfo {author} {\bibfnamefont {L.}~\bibnamefont
  {Zhu}}, \bibinfo {author} {\bibfnamefont {J.}~\bibnamefont {Su}}, \bibinfo
  {author} {\bibfnamefont {C.}~\bibnamefont {Li}}, \ and\ \bibinfo {author}
  {\bibfnamefont {F.-S.}\ \bibnamefont {Zhang}},\ }\bibinfo {title} {How to
  approach the island of stability: Reactions using multinucleon transfer or
  radioactive neutron-rich beams?},\ \href {\doibase
  10.1016/j.physletb.2022.137113} {\bibfield  {journal} {\bibinfo  {journal}
  {Physics Letters B}\ }\textbf {\bibinfo {volume} {829}},\ \bibinfo {pages}
  {137113} (\bibinfo {year} {2022})}\BibitemShut {NoStop}%
\bibitem [{\citenamefont {Liao}\ \emph {et~al.}(2023)\citenamefont {Liao},
  \citenamefont {Zhu}, \citenamefont {Su},\ and\ \citenamefont
  {Li}}]{Liao2023}%
  \BibitemOpen
  \bibfield  {author} {\bibinfo {author} {\bibfnamefont {Z.}~\bibnamefont
  {Liao}}, \bibinfo {author} {\bibfnamefont {L.}~\bibnamefont {Zhu}}, \bibinfo
  {author} {\bibfnamefont {J.}~\bibnamefont {Su}}, \ and\ \bibinfo {author}
  {\bibfnamefont {C.}~\bibnamefont {Li}},\ }\bibinfo {title} {Dynamics of
  charge equilibration and effects on producing neutron-rich isotopes around
  $N=126$ in multinucleon transfer reactions},\ \href {\doibase
  10.1103/PhysRevC.107.014614} {\bibfield  {journal} {\bibinfo  {journal}
  {Phys. Rev. C}\ }\textbf {\bibinfo {volume} {107}},\ \bibinfo {pages}
  {014614} (\bibinfo {year} {2023})}\BibitemShut {NoStop}%
\bibitem [{\citenamefont {Wolschin}\ and\ \citenamefont
  {Nörenberg}(1978)}]{Wolschin1978}%
  \BibitemOpen
  \bibfield  {author} {\bibinfo {author} {\bibfnamefont {G.}~\bibnamefont
  {Wolschin}}\ and\ \bibinfo {author} {\bibfnamefont {W.}~\bibnamefont
  {Nörenberg}},\ }\bibinfo {title} {Analysis of relaxation phenomena in
  heavy-ion collisions},\ \href {\doibase 10.1007/BF01411331} {\bibfield
  {journal} {\bibinfo  {journal} {Zeitschrift für Physik A}\ }\textbf
  {\bibinfo {volume} {284}},\ \bibinfo {pages} {209} (\bibinfo {year}
  {1978})}\BibitemShut {NoStop}%
\bibitem [{\citenamefont {Riedel}\ \emph {et~al.}(1979)\citenamefont {Riedel},
  \citenamefont {Wolschin},\ and\ \citenamefont {Nörenberg}}]{Riedel1979}%
  \BibitemOpen
  \bibfield  {author} {\bibinfo {author} {\bibfnamefont {C.}~\bibnamefont
  {Riedel}}, \bibinfo {author} {\bibfnamefont {G.}~\bibnamefont {Wolschin}}, \
  and\ \bibinfo {author} {\bibfnamefont {W.}~\bibnamefont {Nörenberg}},\
  }\bibinfo {title} {Relaxation times in dissipative heavy-ion collisions},\
  \href {\doibase 10.1007/BF01408479} {\bibfield  {journal} {\bibinfo
  {journal} {Zeitschrift für Physik A: Atoms and Nuclei}\ }\textbf {\bibinfo
  {volume} {290}},\ \bibinfo {pages} {47} (\bibinfo {year} {1979})}\BibitemShut
  {NoStop}%
\bibitem [{\citenamefont {Tōke}\ \emph {et~al.}(1985)\citenamefont {Tōke},
  \citenamefont {Bock}, \citenamefont {Dai}, \citenamefont {Gobbi},
  \citenamefont {Gralla}, \citenamefont {Hildenbrand}, \citenamefont
  {Kuzminski}, \citenamefont {Müller}, \citenamefont {Olmi}, \citenamefont
  {Stelzer}, \citenamefont {Back},\ and\ \citenamefont
  {Bjørnholm}}]{Toke1985}%
  \BibitemOpen
  \bibfield  {author} {\bibinfo {author} {\bibfnamefont {J.}~\bibnamefont
  {Tōke}}, \bibinfo {author} {\bibfnamefont {R.}~\bibnamefont {Bock}},
  \bibinfo {author} {\bibfnamefont {G.}~\bibnamefont {Dai}}, \bibinfo {author}
  {\bibfnamefont {A.}~\bibnamefont {Gobbi}}, \bibinfo {author} {\bibfnamefont
  {S.}~\bibnamefont {Gralla}}, \bibinfo {author} {\bibfnamefont
  {K.}~\bibnamefont {Hildenbrand}}, \bibinfo {author} {\bibfnamefont
  {J.}~\bibnamefont {Kuzminski}}, \bibinfo {author} {\bibfnamefont
  {W.}~\bibnamefont {Müller}}, \bibinfo {author} {\bibfnamefont
  {A.}~\bibnamefont {Olmi}}, \bibinfo {author} {\bibfnamefont {H.}~\bibnamefont
  {Stelzer}}, \bibinfo {author} {\bibfnamefont {B.}~\bibnamefont {Back}}, \
  and\ \bibinfo {author} {\bibfnamefont {S.}~\bibnamefont {Bjørnholm}},\
  }\bibinfo {title} {Quasi-fission — The mass-drift mode in heavy-ion
  reactions},\ \href {\doibase 10.1016/0375-9474(85)90344-6} {\bibfield
  {journal} {\bibinfo  {journal} {Nuclear Physics A}\ }\textbf {\bibinfo
  {volume} {440}},\ \bibinfo {pages} {327} (\bibinfo {year}
  {1985})}\BibitemShut {NoStop}%
\bibitem [{\citenamefont {Li}\ and\ \citenamefont {Wolschin}(1983)}]{Li1983}%
  \BibitemOpen
  \bibfield  {author} {\bibinfo {author} {\bibfnamefont {J.~Q.}\ \bibnamefont
  {Li}}\ and\ \bibinfo {author} {\bibfnamefont {G.}~\bibnamefont {Wolschin}},\
  }\bibinfo {title} {Distribution of the dissipated angular momentum in
  heavy-ion collisions},\ \href {\doibase 10.1103/PhysRevC.27.590} {\bibfield
  {journal} {\bibinfo  {journal} {Physical Review C}\ }\textbf {\bibinfo
  {volume} {27}},\ \bibinfo {pages} {590} (\bibinfo {year} {1983})}\BibitemShut
  {NoStop}%
\bibitem [{\citenamefont {Shen}\ \emph {et~al.}(1987)\citenamefont {Shen},
  \citenamefont {Albinski}, \citenamefont {Gobbi}, \citenamefont {Gralla},
  \citenamefont {Hildenbrand}, \citenamefont {Herrmann}, \citenamefont
  {Kuzminski}, \citenamefont {Müller}, \citenamefont {Stelzer}, \citenamefont
  {Tke}, \citenamefont {Back}, \citenamefont {Bjrnholm},\ and\ \citenamefont
  {Srensen}}]{Shen1987}%
  \BibitemOpen
  \bibfield  {author} {\bibinfo {author} {\bibfnamefont {W.~Q.}\ \bibnamefont
  {Shen}}, \bibinfo {author} {\bibfnamefont {J.}~\bibnamefont {Albinski}},
  \bibinfo {author} {\bibfnamefont {A.}~\bibnamefont {Gobbi}}, \bibinfo
  {author} {\bibfnamefont {S.}~\bibnamefont {Gralla}}, \bibinfo {author}
  {\bibfnamefont {K.~D.}\ \bibnamefont {Hildenbrand}}, \bibinfo {author}
  {\bibfnamefont {N.}~\bibnamefont {Herrmann}}, \bibinfo {author}
  {\bibfnamefont {J.}~\bibnamefont {Kuzminski}}, \bibinfo {author}
  {\bibfnamefont {W.~F.~J.}\ \bibnamefont {Müller}}, \bibinfo {author}
  {\bibfnamefont {H.}~\bibnamefont {Stelzer}}, \bibinfo {author} {\bibfnamefont
  {J.}~\bibnamefont {Tke}}, \bibinfo {author} {\bibfnamefont {B.~B.}\
  \bibnamefont {Back}}, \bibinfo {author} {\bibfnamefont {S.}~\bibnamefont
  {Bjrnholm}}, \ and\ \bibinfo {author} {\bibfnamefont {S.~P.}\ \bibnamefont
  {Srensen}},\ }\bibinfo {title} {Fission and quasifission in U-induced
  reactions},\ \href {\doibase 10.1103/PhysRevC.36.115} {\bibfield  {journal}
  {\bibinfo  {journal} {Physical Review C}\ }\textbf {\bibinfo {volume} {36}},\
  \bibinfo {pages} {115} (\bibinfo {year} {1987})}\BibitemShut {NoStop}%
\end{thebibliography}%


%



\end{document}